%


\input epsf

\magnification=1200
\pageno=0
\hsize=5.6in
\vsize=7.5in
\voffset=0.2in
\hoffset= -0.2in

\font\titulo=cmb10 scaled 1440


\def\normal{{\bigcirc \!\hskip -5pt n}}

\def\0barra{{\rm O} \!\hskip -3.7pt {\rm l} }

\def\1barra{1\! \hskip -1.1pt {\rm l}}


\baselineskip=18pt

\centerline{\titulo Analytical Results for the Grand Canonical} \par

\centerline{\titulo  Partition Function for Unidimensional Hubbard Model,} 
\par

\centerline{\titulo  Up to Order $\beta^5$.}

\baselineskip=12pt
\vskip 0.3cm

\centerline {I.C. Charret\footnote{$^1$}{E--mail: iraziet@if.uff.br}, 
} \par
\centerline {\it Departamento de F\'\i sica} \par
\centerline {\it Instituto de Ci\^encias Exatas} \par
\centerline {\it Universidade Federal de Minas Gerais} \par
\centerline{\it Campus da Pampulha} \par
\centerline {\it Belo Horizonte, M.G., 31270--901 $\quad$ BRAZIL} \par 

\vskip 0.2cm 

\centerline{ E.V. Corr\^ea Silva\footnote{$^2$}{E--mail:
ecorrea@cbpfsu1.cat.cbpf.br }} \par
\centerline{\it Centro Brasileiro de Pesquisas F\'{\i}sicas} \par
\centerline{\it R. Dr. Xavier Sigaud n.$\!\!^{\rm o}$ 150} \par
\centerline{\it Rio de Janeiro, R.J., 22290-180 $\quad$ BRAZIL } \par 

\vskip 0.2cm

\centerline{  S.M.  de Souza\footnote{$^3$}{E--mail: smartins@if.uff.br},
} \par
\centerline {\it Departamento de Ci\^encias Exatas} \par
\centerline {\it Universidade Federal de Lavras} \par
\centerline {\it C.P.: 37} \par
\centerline {\it Lavras, M.G., 37200--000 $\quad$ BRAZIL} \par 

\vskip 0.2cm

\centerline{ M.T. Thomaz\footnote{$^4$}{Corresponding author: 
Dr. Maria Teresa Thomaz;
 R. Domingos S\'avio Nogueira Saad n.$\!\!^{\rm o}$ 120  apto 404, 
 Niter\'oi, R.J., 24210--340, BRAZIL
 --Phone/Fax: (21) 620--6735; \hfill \break E--mail: mtt@if.uff.br }
} \par
\centerline {\it Instituto de  F\'\i sica} \par
\centerline {\it Universidade Federal Fluminense} \par
\centerline {\it Av. Gal. Milton Tavares de Souza s/n.$\!\!^o$} \par
\centerline{\it Campus da Praia Vermelha} \par
\centerline {\it Niter\'oi, R.J., 24210--310 $\quad$ BRAZIL} \par

\baselineskip=14pt

\vskip 0.3cm

\centerline {\bf Abstract} \par

\vskip 0.2cm

We  calculate the exact analytical coefficients
of the $\beta$--expansion of the grand canonical partition function of
the unidimensional Hubbard model up to order $\beta^5$, using
an alternative method, based on properties of the Grassmann algebra. 
The results derived are non--perturbative  and no
restrictions on the set of parameters that characterize the model are
required. By applying this  method
we obtain analytical results for the themodynamical
quantities, in the high--temperature limits, for arbitrary density
of electrons in the unidimensional chain.  \par

\vfill




\eject

\baselineskip=18pt

\noindent {\bf 1. Introduction} \par

Working with fermionic variables seems discouraging, for their
non--commuting nature, in a certain way, compells one to a higher
degree of care than that required by commutative variables.
Nevertheless, Grassmann algebra properties justify their 
use in many circumstances [1]. With that in mind, we have recently
described an alternative method to calculate the terms of the
$\beta$-- expansion of the grand canonical partition function of
periodic unidimensional self--interacting fermionic models, in which
Grassmann algebra properties play a central role. No auxiliary fields
are needed, and we express our results in terms of matrices with
commuting elements. The general approach for a periodic unidimensional
fermionic model has been applied to the unidimensional Hubbard model up to
order $\beta^3$ [2]. An important point about the method
developed in reference [2] is that even though the unidimensional 
Hubbard model has exact solutions [3], the analytical expressions 
are only known in the half-- filled case. 
This drawback hinders the analytical  evaluation of the
partition function for the model from the knowledge of its energy
spectrum when we are not considering the half--filling case. 
Takahashi[4] derived a closed expression for the grand
canonical partition function of the unidimensional Hubbard model, but
besides  requiring  the formulation of some additional hypotesis, a
simple closed expression for the grand canonical partition function
 is only  obtained in the
strong coupling limit. The literature offers many examples of high
temperature expansions of the grand canonical partition function for
the Hubbard model in different space dimensions, some of them up to
order $\beta^{9}$ [5]. However, all these works referer to either 
some approximation in which one of the characteristic constants of the
model has to be much bigger than the other, or some consideration
based on important numerical analysis. \par

Our results are {\it analytical} and do not rest upon any additional
hypothesis on the constants that characterize the model. We should
point out that we do {\it not} perform a perturbative expansion 
valid in the high temperature limit [5], 
but a $\beta$--expansion of the grand canonical partition
function [6] where we calculate the exact analytical coefficients of the
terms up to order $\beta^5$ for any density of electrons 
in the unidimensional chain. \par

In the present paper, we will apply the approach developed in
reference [2] to get the exact terms at orders $\beta^4$ and $\beta^5$
of the grand canonical partition function of the unidimensional
Hubbard model. In section 2 we present a review of the results of
reference [2]. In section 3, we write down the unidimensional Hubbard
model and the Grassmann functions necessary in the calculations
that follow. In section 4 we obtain the
 coefficients at orders $\beta^4$ and $\beta^5$
of the $\beta$--expansion of the grand canonical partition function of
the unidimensional Hubbard model. We use the grand potencial derived
up to order $\beta^4$ to calculate some physical quantities. A certain
property of the multivariable Grassmann integrals --- namely, its
factorization --- opens the way to extending the method of reference
[2] to orders higher than $\beta^3$. This factorization property is
presented in Appendix A through one example. In Appendix B we
introduce a graphical notation that greatly simplifies the
calculations. In Appendix C, we present a table of the necessary
multivariable Grassmann integrals. Section 5 is dedicated to our
conclusions. \par

\vskip 1cm

\noindent {\bf 2. Review of Previous Results [2]}  \par

The grand canonical partition function of any system is given by: \par

$$
{\cal Z}(\beta; \mu) = {\rm Tr}( e^{- \beta {\bf K}}),  \eqno(2.1)
$$

\noindent where $\beta = {1 \over kT}$, {\it k} is the Boltzmann constant,
 {\it T} is the absolute temperature, and  \par

$$ {\bf K} = {\bf H} - \mu {\bf N}, \eqno (2.1a)
$$

\noindent {\bf H} is the hamiltonian of the system, $\mu$ is the
chemical potential and {\bf N} is the total number of particles
operator. \par

In the high temperature limit, $\beta \ll 1$, ${\cal Z}(\beta; \mu)$
has the expansion   \par

$$
{\cal Z}(\beta, \mu) = {\rm Tr}[ 1\!\hskip -1pt{\rm I}  - \beta {\bf K}] +
\;\;    \sum_{ n=2}^{\infty} {(-\beta)^n \over n!}\;\; {\rm Tr}[ {\bf K}^n],
      \eqno(2.2)  $$

\noindent that we call the $\beta$--expansion of the grand canonical 
partition function. \par

For any self--interacting fermionic quantum system, in reference [2]
we used the Grassmann algebra to show that   \par

\vskip -0.7cm

$$\eqalignno{ {\rm Tr}[{\bf K}^n] &= \int \prod_{I=1}^{2nN} \, d\eta_I
d\bar{\eta}_I \;\; e^{\sum\limits_{I,J= 1}^{2nN} \bar{\eta}_I\; A_{I
J} \; \eta_J} \times & \cr
%
%
 & \hskip -1cm \times
{\cal K}^{\normal} (\bar{\eta}, \eta; \nu=0)\; {\cal K}^{\normal}
(\bar{\eta}, \eta; \nu=1) \; \cdots \; {\cal K}^{\normal} (\bar{\eta},
\eta; \nu=n-1), & (2.3) }$$

\noindent where the matrix {\bf A} is given by,  \par

$$ {\bf A} = \pmatrix { {\bf A}^{\uparrow \uparrow} & \0barra \cr
 & &   \cr
   \0barra & {\bf A}^{\downarrow \downarrow} \cr }, \eqno (2.3a) $$

\noindent and  \par

$$ {\bf A}^{\uparrow \uparrow} = {\bf A}^{\downarrow \downarrow} =
\pmatrix { \1barra_{N\times N} & - \1barra_{N\times N} &
\0barra_{N\times N} & \cdots & \0barra_{N\times N} \cr & & & & \cr
%
%
\0barra_{N\times N} & \1barra_{N\times N}&
- \1barra_{N\times N} & \cdots & \0barra_{N\times N}\cr
%
%
& & & & \cr \vdots & & & & \vdots \cr
%
%
 \1barra_{N\times N} & \0barra_{N\times N} &
\0barra_{N\times N} & \cdots & \1barra_{N\times N} \cr}. \eqno(2.3b)
$$

\noindent Each matrix  ${\bf A}^{\sigma \sigma}$ has dimension
$ nN \times nN$, $\1barra_{N\times N}$ been the identity matrix in dimension
$N\times N$ and $\0barra_{N\times N}$  the null matrix in this
dimension. {\it N} is the number of space sites and {\it n} is the
power of the $\beta$ term.
The non--null elements of ${\bf A}^{\sigma \sigma}$ ,
$\sigma= \uparrow$ and $\sigma= \downarrow$, are    \par

$$ A_{I J}^{\sigma \sigma} = \cases{ a_{II} = 1, & $I= 1, 2, \cdots,
nN $ \cr & \cr a_{I, I+N} = -1, & $I= 1, 2, \cdots, (n-2)N $ \cr & \cr
a_{(n-1)N +I, I} = 1, & $ I= 1, 2, \cdots, N. $ \cr} \eqno(2.3c) $$

\vskip 0.5cm

\noindent The Grassmann function
${\cal K}^{\normal} (\bar{\eta}, \eta)$ is
the kernel of the fermionic operator {\bf K} in the normal order [7]. In
writing down eq.(2.3), we have used a particular mapping for the Grassmann
generators [2], that greatly  simplify our calculations:  \par

\vskip -0.5cm

$$\eqalignno{ \eta_\uparrow (x_l, \tau_\nu) &\equiv \eta_{\nu N + l} &
(2.4a) \cr
\noalign {\hbox{ and}}
 \eta_\downarrow (x_l, \tau_\nu)
&\equiv \eta_{(n+\nu) N + l}, & (2.4b) \cr }$$

\noindent where $ l= 1, 2, \cdots, N$, and $ \nu= 0, 1, \cdots, n-1$. 
The mappings (2.4a--b) can be summarized as [8]: \par

$$ \eta_\sigma (x_l, \tau_\nu) \equiv \eta_{[{(1-\sigma)\over 2} n +
\nu]N + l}\hskip 4pt. \eqno (2.4c) $$

\noindent The generators $\bar{\eta}_\sigma (x_l, \tau_\nu)$
have an equivalent mapping.  \par

In reference [9] we showed that the Grassmann integrals (2.3) can be written
as co--factors of the matrix ${\bf A}^{\sigma \sigma}$. 
As we have stated  before,
the calculation of the co--factors of matrix ${\bf A}^{\sigma \sigma}$
gets simpler when we diagonalize it through a similarity transformation
\par

$$ {\bf P}^{-1} {\bf A}^{\sigma\sigma} {\bf P} = {\bf D}, \eqno(2.5)
$$

\noindent where the matrix {\bf D} is, \par

$$ {\bf D} = \pmatrix{ \lambda_1 \1barra_{N\times N} &
\0barra_{N\times N} & \cdots & \0barra_{N\times N} \cr
\0barra_{N\times N} & \lambda_2 \1barra_{N\times N} & \cdots &
\0barra_{N\times N} \cr
 \vdots & & & \vdots\cr
\0barra_{N\times N} & \0barra_{N\times N} & \cdots & \lambda_n
\1barra_{N\times N} \cr }, \eqno(2.5a)$$

\noindent $\lambda_i$, $i= 1, 2, \cdots, n$, are the eigenvalues of
matrices ${\bf A}^{\sigma\sigma}$, $\sigma= \uparrow, \downarrow$. At
the same time, we transform of the anti--commuting variables, \par

$$ \eta^\prime = {\bf P}^{-1} \, \eta \hskip 1cm {\rm and} \hskip 1cm
\bar{\eta}^\prime = \bar{\eta} \,{\bf P}. \eqno(2.5b) $$

The matrices {\bf P} and ${\bf P}^{-1}$ have a block--structure, and in
reference [2] we got the elements of matrix {\bf P} and its inverse for the
particular cases $n=2$ and $n=3$, besides the eigenvalues of matrices
${\bf A}^{\sigma  \sigma}$, $\sigma= \uparrow, \downarrow$.  However,
for any value of {\it n}, we have that   \par

$$ \eqalignno{ p_{\nu \nu^\prime}^{(n)} &= {1 \over \sqrt{n}}
e^{{i \pi \over n} (2\nu^\prime + 1)(\nu +1)},   &  (2.6a)  \cr
\noalign{\hbox{and}}
q_{\nu^\prime \nu}^{(n)} &= {1 \over \sqrt{n}}
e^{-{i \pi \over n} (2\nu^\prime + 1)(\nu +1)},   &  (2.6b)  \cr
} $$

\noindent with $\nu, \; \nu^\prime = 0, 1, \cdots, n-1$, and  \par

$$ {\bf P} = \pmatrix{ p_{0 0}^{(n)}\1barra_{N\times N} &
 \cdots & p_{0, n-1}^{(n)} \1barra_{N\times N} \cr
 \vdots & & \vdots\cr
p_{n-1, 0}^{(n)}\1barra_{N\times N} &  \cdots & p_{n-1, n-1}^{(n)}
\1barra_{N\times N} \cr } \eqno(2.6c)$$

\noindent and    \par

$$ {\bf P}^{-1} = \pmatrix{ q_{0 0}^{(n)}\1barra_{N\times N} &
 \cdots & q_{0, n-1}^{(n)} \1barra_{N\times N} \cr
 \vdots &  & \vdots\cr
q_{n-1, 0}^{(n)}\1barra_{N\times N} &  \cdots & q_{n-1, n-1}^{(n)}
\1barra_{N\times N} \cr }. \eqno(2.6d)$$

\vskip 0.5cm

\noindent The diagonal elements of matrix {\bf D} are:  \par

\vskip -0.5cm

$$
\lambda_\nu^{(n)} = 1 - e^{{i\pi \over n} (2\nu +1)},
\hskip 1cm \nu = 0, 1, \cdots, n-1.   \eqno(2.6e)
$$

\noindent where the eigenvalues are {\it N}--fold degenerated,
$N$ being the number of space sites.
 Due to space translation symmetry, we should note that the
elements $p_{\nu \nu^\prime}^{(n)}$ and $q_{\nu \nu^\prime}^{(n)}$ do not
carry any space site index.   \par

Once the matrix {\bf A} has a block--structure as depicted
in eq.(2.3), the integrals (2.3) are
equal to the product of the integral in the sector
$\sigma\sigma = \uparrow\uparrow$ times the integral in the sector
$\sigma\sigma = \downarrow\downarrow$. The Grassmann integrals in
the  sector $\uparrow\uparrow$ have the form:

$$M(L,K) = \int \prod_{i = 1}^{\rm nN} d\eta_i d\bar\eta_i
 \;
\bar\eta_{l_1} \eta_{k_1} \cdots \bar\eta_{l_m} \eta_{k_m} \hskip 3pt
 e^{ \sum\limits_{ i, j = 1}^{\rm nN} \bar\eta_i
A_{i j}^{\uparrow\uparrow} \eta_j},
\eqno(2.7) $$

\noindent with $ L= \{ l_1, \cdots, l_m\}$ and $K= \{k_1, \cdots,
k_m\}$. The products $\bar{\eta}\eta$ are ordered in such a way that
$ l_1< l_2< \cdots< l_m$ and $ k_1< k_2< \cdots< k_m$. From reference [9],
the result of this type of integrals is equal to: \par

$$ M(L,K) = (-1)^{(l_1 + l_2+ \cdots+ \l_m) + (k_1+ k_2+ \cdots+ k_m)}
A(L,K), \eqno(2.7a) $$

\noindent where $A(L,K)$ is the determinant of the matrix obtained
from matrix {\bf A} by deleting the lines $\{ l_1, \cdots, l_m\}$ and
the columns: $\{k_1, \cdots, k_m\}$.  The Grassmann integrals to be 
calculated in sector $\downarrow\downarrow$ are of the same 
type as that of eq.(2.7).\par

After the similarity  transformation (2.5) and the change of variables
(2.5b), in a schematic way, the integral (2.7) becomes   \par

$$ \eqalignno{ M(L,K) &= \int \prod_{i = 1}^{\rm nN} d\eta_i
d\bar\eta_i
 \; (\bar\eta {\bf P}^{-1})_{l_1} ({\bf P}\eta)_{k_1}
\cdots (\bar\eta {\bf P}^{-1})_{l_m} ({\bf P} \eta)_{k_m}
	\hskip 6pt
 e^{ \sum\limits_{ i, j = 1}^{\rm nN} \bar\eta_i D_{i
j} \eta_j}, &    \cr
%
%
     &         & (2.8) \cr }$$

\noindent where {\bf D} is a diagonal matrix whose entries are
given by eq.(2.6e).   \par

We should point out that the relations (2.6a)--(2.6e) are valid  for
any unidimensional self--interacting fermionic model with space
translation symmetry.   \par

We have a large number of integrals that contribute to eq.(2.3). For a
discussion on some useful symmetries and their use in the reduction of
the number of contributing integrals, the reader is referred to [2].
\par

\vskip 1cm

\noindent {\bf 3. Unidimensional Hubbard Model}   \par

The hamiltonian that describes the Hubbard model in one space
 dimension is [10]:

$${\bf H} = \sum_{ {i=1}\atop{j=1}}^{\rm N} \sum_{\sigma= -1, 1}
t_{ij} {\bf a}_{i\sigma}^{\dagger}{\bf a}_{j\sigma} + U \sum_
{i=1}^{\rm N} {\bf a}_{i\uparrow}^\dagger{\bf a}_{i\uparrow} {\bf
a}_{i\downarrow}^\dagger{\bf a}_{i\downarrow} + \lambda_B \sum_
{i=1}^{\rm N} \sum_{\sigma=-1, 1} \sigma {\bf a}_{i\sigma}^\dagger{\bf
a}_{i\sigma} \eqno (3.1) $$

\noindent where ${\bf a}_{i\sigma}^\dagger$ is the creation operator
of an electron in site {\it i} with spin  $\sigma$ and
 ${\bf a}_{i\sigma}$  is the destruction
operator of an electron in site {\it i} with spin $\sigma$.  The first
term on the r.h.s. of eq.(3.1) is the kinetic energy operator. All
diagonal elements of $t_{ij}$ are equal, $t_{ii} = {\rm E_0}$, 
the only non--null off--diagonal terms
are $t_{i, i-1}=t_{i, i+1}= t$, where $ i=1, 2, \dots , {\rm N}$, and
they contribute to the hopping term. {\it U} is the strength of the
interaction between the electrons in the same site but with different
spins. We have defined $\lambda_B= -{1\over 2} g\mu_B B$, where {\it
g} is the Land\'e's factor, $\mu_B$ is the Bohr's magneton and
$B$ is the constant external magnetic field in the $\hat{z}$
direction. \par

The periodic boundary condition in space is implemented by imposing
that ${\bf a}_{0 \sigma} \equiv {\bf a}_{N \sigma}$ and ${\bf a}_{N+1,
\sigma} \equiv {\bf a}_{1 \sigma}$. Therefore, the hopping terms $t_{1
0} {\bf a}_{1 \sigma}^\dagger {\bf a}_{0 \sigma}$ and $t_{N, N+1} {\bf
a}_{N \sigma}^\dagger {\bf a}_{N+1, \sigma}$ become $t_{1 N} {\bf
a}_{1 \sigma}^\dagger {\bf a}_{N\sigma}$ and $t_{N, 1} {\bf a}_{N
\sigma}^\dagger {\bf a}_{1 \sigma}$ respectively. We point out that
the hamiltonian (3.1) is already in normal order. \par

The kernel of the operator {\bf K} (eq.(2.1a))  for the unidimensional
Hubbard model on a lattice  with {\it N} space sites, written in
terms of the generators $\bar{\eta}_I$ and $\eta_J$, is equal to   \par

$$\eqalignno{ {\cal K}^{\normal} &(\bar{\eta},\eta;\nu)) =
\sum_{l=1}^N \sum_{\sigma =\pm 1} (E_0 + \sigma \lambda_B - \mu) \;
\bar{\eta}_{[{(1-\sigma)\over 2}n +\nu]N + l}\;
\eta_{[{(1-\sigma)\over 2}n +\nu]N + l} + & \cr
%
%
& + \sum_{l=1}^N \sum_{\sigma =\pm 1} t [
\bar{\eta}_{[{(1-\sigma)\over 2}n +\nu]N + l}\;
\eta_{[{(1-\sigma)\over 2}n +\nu]N + l+1}+
\bar{\eta}_{[{(1-\sigma)\over 2}n +\nu]N + l}\;
\eta_{[{(1-\sigma)\over 2}n +\nu]N + l-1}] + & \cr
%
%
 & + \sum_{l=1}^N U \bar{\eta}_{(n+\nu)N + l}\;\;
\eta_{(n+\nu)N + l}\;\; \bar{\eta}_{\nu N + l} \; \; \eta_{\nu N + l},
& (3.2a) \cr }$$

\noindent with the  periodic spatial boundary  condition:   \par

\vskip -0.7cm

$$ \eqalignno{ t\; \bar{\eta}_{[{(1-\sigma)\over 2} n + \nu]N + N}
\; \eta_{[{(1-\sigma)\over 2} n + \nu]N + N +1}
& \equiv t \; \bar{\eta}_{[{(1-\sigma)\over 2} n +\nu]N + N} \;
\eta_{[{(1-\sigma)\over 2} n + \nu]N + 1} & (3.2b)  \cr
\noalign{\hbox{and}}
t \; \bar{\eta} _{[{(1-\sigma)\over 2} n + \nu]N + 1}  \;
\eta_{[{(1-\sigma)\over 2} n + \nu]N } & \equiv
t \; \bar{\eta}_{[{(1-\sigma)\over 2} n +\nu]N + 1} \;
\eta_{[{(1-\sigma)\over 2} n +\nu]N + N}\;\; , &
(3.2c) \cr } $$

\noindent and the anti--periodic boundary condition in $\nu$: \par

$$ \eta_{[{(1-\sigma)\over 2} n +  n]N + l} =
- \eta_{[{(1-\sigma)\over 2} n ]N + l}\;\;, \eqno
(3.2d) $$

\noindent for $l= 1, 2, \cdots, N$, and $\sigma= \uparrow,
\downarrow$.  We have used the mapping (2.4c) to write 
the previous expressions. \par

In order to write down the terms that
contribute to $Tr[{\bf K}^4]$ and $Tr[{\bf K}^5]$
in a simplified way , we define:  \par

$$ \eqalignno{ {\cal E} (\bar{\eta}, \eta; \nu; \sigma) &\equiv
\sum\limits_{l=1}^N \bar{\eta}_{[{(1-\sigma)\over 2}n +\nu]N + l}\;
\eta_{[{(1-\sigma)\over 2}n +\nu]N + l}\;\;; & (3.3a) \cr
%
%
{\cal T}^{-} (\bar{\eta}, \eta; \nu; \sigma) & \equiv
\sum\limits_{l=1}^N \bar{\eta}_{[{(1-\sigma)\over 2}n +\nu]N + l}\;
\eta_{[{(1-\sigma)\over 2}n +\nu]N + l+1} \;\; ; & (3.3b)\cr
\noalign{\hbox{and}}
%
%
 {\cal T}^{+} (\bar{\eta}, \eta; \nu; \sigma) &\equiv
\sum\limits_{l=1}^N \bar{\eta}_{[{(1-\sigma)\over 2}n +\nu]N + l}\;
\eta_{[{(1-\sigma)\over 2}n +\nu]N + l-1}\;\; . & (3.3c)\cr
}$$

\noindent We also define \par

$$\eqalignno{ {\cal E} (\bar{\eta}, \eta; \nu) &\equiv
\sum\limits_{\sigma=\pm 1} E(\sigma) {\cal E}(\bar{\eta}, \eta; \nu;
\sigma), & (3.4a) \cr
%
%
 {\cal T}^{-} (\bar{\eta},
\eta; \nu) &\equiv \sum\limits_{\sigma = \pm 1} t\, {\cal
T}^{-}(\bar{\eta}, \eta; \nu; \sigma), & (3.4b) \cr
%
%
 {\cal T}^{+} (\bar{\eta},
\eta; \nu) &\equiv \sum\limits_{\sigma =\pm 1} t \,{\cal
T}^{+}(\bar{\eta}, \eta; \nu; \sigma), & (3.4c) \cr
\noalign{\hbox{and}}
%
%
 {\cal U}(\bar{\eta}, \eta; \nu)
&\equiv \sum\limits_{l=1}^N \bar{\eta}_{(n+\nu)N + l}\;\;
\eta_{(n+\nu)N + l}\;\; \bar{\eta}_{\nu N + l} \; \; \eta_{\nu N + l},
& (3.4d) \cr   }$$

\noindent where $E(\sigma) \equiv E_0 - \sigma\lambda_B - \mu$. The
term ${\cal E} (\bar{\eta}, \eta; \nu)$ represents the diagonal part of 
the kinetic energy , ${\cal T}^{-} (\bar{\eta}, \eta; \nu)$
 and $ {\cal T}^{+}
(\bar{\eta}, \eta; \nu)$ are the hopping terms and ${\cal U}
(\bar{\eta}, \eta; \nu)$ is the fermionic interaction term. \par

For the unidimensional Hubbard model, the Grassmannian function
${\cal K}^{\normal} (\bar{\eta},\eta;\nu)$ is written as \par

$${\cal K}^{\normal} (\bar{\eta}, \eta; \nu)= {\cal E} (\bar{\eta},
\eta; \nu)+ {\cal T}^{-} (\bar{\eta}, \eta; \nu)+ {\cal T}^{+}
(\bar{\eta}, \eta; \nu) + {\cal U} (\bar{\eta}, \eta; \nu).
\eqno(3.5) $$

\vskip 1cm

\noindent {\bf 4. The Exact Coefficients of the $\beta$--Expansion
of the Grand Canonical Partition Function for the Unidimensional
Hubbard Model}     \par

In eq.(2.2) we have the $\beta$--expansion of the grand canonical
 partition function for any quantum system. For the particular case
of self--interacting unidimensional fermionic models, the expression
of $Tr[{\bf K}^n]$ is given by eq.(2.3).   \par

In reference [2], we calculated the exact coefficients of the terms
$\beta^2$ and $\beta^3$ of the expression (2.2) for the unidimensional
Hubbard model for arbitrary values of the constants $E_0$, $t$, $U$
and $\mu$, that characterize the model, and for any value of the
constant external magnetic field $B$.

The evaluation of integrals has been performed by a number of
procedures (computer programs) developed by the authors in the
symbolic language Maple V.3, that consist in the computational
implementation of the method described in reference [2]. We have
called this package of procedures {\tt GINT}.

The method applied in [2] greatly simplifies the calculations made in
reference [11], but memory utilization problems have appeared as we
tried to go beyond $n>3$. Luckily, the factorization property of
multivariable integrals of type (2.8) allowed us to optimize the
performance of the package. In Appendix A we consider one typical
Grassmann integral to exemplify the factorization of the sub--graphs.
\par

The procedure {\tt perm } is one of the procedures contained in the
package, being a useful tool to calculate the independent non--null
terms [12]  that contribute to
$Tr[{\bf K}^4]$ and $Tr[{\bf K}^5]$. In this procedure are implemented
the symmetries discussed in reference [2]. The procedure {\tt gint}
has been used to calculate the multivariable Grassmann integrals,
taking into account the factorization into sub--graphs. The package 
can be downloaded through {\it ftp} from the site
{\it http:/www.if.uff.br}. \par

\vskip 1cm

\noindent {\bf 4.1. Calculation of Tr[${\bf K}^4$]}   \par

For the case $n=4$, we get from eq.(2.3) that  \par

\vskip -0.7cm

$$\eqalignno{ {\rm Tr}[{\bf K}^4] &= \int \prod_{I=1}^{8N} \, d\eta_I
d\bar{\eta}_I \;\; e^{\sum\limits_{I,J= 1}^{8N} \bar{\eta}_I\; A_{I
J} \; \eta_J} \times & \cr
%
%
 & \hskip -1cm \times
{\cal K}^{\normal} (\bar{\eta}, \eta; \nu=0)\; {\cal K}^{\normal}
(\bar{\eta}, \eta; \nu=1) \;
{\cal K}^{\normal} (\bar{\eta}, \eta; \nu=2)\;
 {\cal K}^{\normal} (\bar{\eta}, \eta; \nu=3). & (4.1.1)
 }$$

\noindent The expressions of ${\bf A}^{\sigma \sigma}$,
 $p_{\nu \nu^\prime}^{(4)}$, $q_{\nu \nu^\prime}^{(4)}$ and
$\lambda_\nu^{(4)}$ are obtained  from eqs.(2.3b), (2.6a), (2.6b) and
 (2.6e).   \par

From eq.(3.5), for the unidimensional Hubbard model, we have that
the Grassmann function ${\cal K}^{\normal} (\bar{\eta}, \eta; \nu)$
is equal to   \par

$${\cal K}^{\normal} (\bar{\eta}, \eta; \nu)= {\cal E} (\bar{\eta},
\eta; \nu)+ {\cal T}^{-} (\bar{\eta}, \eta; \nu)+ {\cal T}^{+}
(\bar{\eta}, \eta; \nu) + {\cal U} (\bar{\eta}, \eta; \nu).
\eqno(4.1.1a) $$

We defined a simplified notation,   \par

$$\eqalignno{ < {\cal O}_1 (\nu_1) \cdots {\cal O}_m (\nu_m) > &
\equiv \int \prod_{I=1}^{2nN} \, d\eta_I d\bar{\eta}_I \;\;
e^{\sum\limits_{I,J= 1}^{2nN} \bar{\eta}_I\; A_{I J} \; \eta_J} \;\;
\times & \cr
 & \hskip 1cm \times \;\; {\cal O}_1
(\bar{\eta},\eta;\nu_1) \cdots {\cal O}_m (\bar{\eta},\eta;\nu_m) &
(4.1.2a) \cr
 \noalign{\hbox{and}}
%
%
 < {\cal O}_1 (\sigma, \nu_1) \cdots {\cal O}_m (\sigma,\nu_m) >
&\equiv \int
\prod_{I=(1-\sigma)nN+1}^{(3-\sigma)nN}\, d\eta_I d\bar{\eta}_I \;\;
e^{\sum\limits_{I,J= 1}^{2nN} \bar{\eta}_{(1-\sigma)nN+I} \; A_{I J}
\; \eta_{(1-\sigma)nN+J}} \;\; \times & \cr
 & \hskip 1cm \times
{\cal O}_1 (\bar{\eta},\eta;\nu_1) \cdots {\cal O}_m
(\bar{\eta},\eta;\nu_m), & (4.1.2b) \cr }$$

\noindent where ${\cal O}_j (\bar{\eta},\eta;\nu_j)$ are Grassmann
functions.   \par

The independent terms that contribute to
$Tr[{\bf K}^4]$ are:  \par

\vskip -0.7cm

$$\eqalignno{
%
Tr[{\bf K}^4] &= <{\cal E}_0, {\cal E}_0, {\cal E}_0,{\cal E}_0> +
4 <{\cal U}, {\cal E}_0, {\cal E}_0, {\cal E}_0> +
2 <{\cal U}, {\cal E}_0, {\cal U}, {\cal E}_0> +
8 < {\cal U},{\cal T}^-,{\cal T}^+, {\cal E}_0> +    &  \cr
%
%
 & + 4 <{\cal U}, {\cal U}, {\cal E}_0, {\cal E}_0> +
4 < {\cal U}, {\cal U}, {\cal U}, {\cal E}_0> +
< {\cal U}, {\cal U}, {\cal U}, {\cal U}> +
4 < {\cal T}^-, {\cal E}_0, {\cal T}^+, {\cal E}_0> +  & \cr
%
%
 & + 8 < {\cal T}^-, {\cal U}, {\cal T}^+, {\cal E}_0> +
4 < {\cal T}^-, {\cal U}, {\cal T}^+, {\cal U}> +
8 < {\cal T}^-, {\cal T}^+, {\cal E}_0, {\cal E}_0> + &   \cr
%
%
 &  + 8 < {\cal T}^-, {\cal T}^+, {\cal U}, {\cal E}_0> +
 8 < {\cal T}^-, {\cal T}^+, {\cal U}, {\cal U}> +
2 < {\cal T}^-, {\cal T}^+, {\cal T}^-, {\cal T}^+> +  &   \cr
%
%
 &  + 4 <{\cal T}^-, {\cal T}^-, {\cal T}^+, {\cal T}^+>.  
   & (4.1.3) \cr
}$$

In order to calculate the terms on the r.h.s. of (4.1.3), 
we need the result of
a set of Grassmann multivariable integrals
that are presented on Appendix C.
We used the procedure {\tt gint} to calculate them. Before using
the procedure {\tt gint} to calculate the terms in eq.(4.1.3) that include
${\cal E}_0$, $ {\cal T}^-$ and ${\cal T}^+$, we need to explicit the
contributions coming from  the sectors $\sigma= \uparrow$ and
$\sigma=\downarrow$. For example, in the term
$< {\cal E}(0), {\cal E}(1), {\cal E}(2), {\cal E}(3) > $ we have
16 terms when we explicitly  write down the spin--sectors. However, the
number of terms is diminished when we use the symmetries
discussed in reference [2] and the fact that
${\bf A}^{\uparrow \uparrow}$ =
${\bf A}^{\downarrow \downarrow}$. The application  of these
symmetries is simplified by using the graphic notation
explained  in Appendix B.    \par

\vskip 0.5cm

By taking into account the results of integrals in Appendix C
 and their contributions to the sum over the space indices  in
each term of $Tr[{\bf K}^4]$ (eq.(4.1.3)), we finally obtain \par

\vskip -0.3cm

$$ \eqalignno{
Tr[{\bf K}^4]  &= N^4 \biggl( U \Delta E^3 + {1\over 256} U^4 +
{3\over 8} U^2 \Delta E^2 + \Delta E^4 + {1\over 16}U^3 \Delta E \biggr) +
 N^3 \biggl( 3 \Delta E^2 \lambda_B^2 + 3 \Delta E^4 +
&    \cr
& \hskip 2cm   + 3 t^2 U \Delta E+
{9 \over 2} U \Delta E^3 + {3 \over 16} U^2 \lambda_B^2 + 6t^2 \Delta E^2 +
{9 \over 128} U^4  + {45 \over 16} U^2 \Delta E^2 +    &   \cr
& + {3 \over 8} t^2 U^2  + {3\over 4} U^3 \Delta E +
{3 \over 2}  U \Delta E \lambda_B^2  \biggr)  +
 N^2 \biggl( {3\over 4} \Delta E^4 + {3\over  16} U^2 \lambda_B^2 +
3 U \Delta E^3  + {51 \over 16} U^2 \Delta E^2 + &    \cr
& + 3 t^2 U \Delta E + {3\over 2} \Delta E^2 \lambda_B^2 +
 3 t^2 \lambda_B^2 + {3 \over 4} \lambda_B^4+
 {9\over 8} t^2 U^2 + 3 t^4 + 3 t^2 \Delta E^2 + {21\over 16} U^3 \Delta E
+ {51 \over 256} U^4 \biggr) +  &  \cr
& - N \biggl( 3 t^2 \lambda_B^2 + {3\over 2} t^4 + 3 t^2 \Delta E^2  +
{1\over 2} U \Delta E^3 + {5\over 4} t^2 U^2 + {3\over 8}  U^2 \lambda_B^2
+ {1\over 4} \Delta E^4  +      &   \cr
&  + 3 t^2 U \Delta E + {3\over 2} \Delta E^2 \lambda_B^2 +
{3\over 8} U^2 \Delta E^2 + {1\over 4} \lambda_B^4 +
{1\over 8} U^3 \Delta E + {3\over 2} U \Delta E \lambda_B^2 +
{3 \over 128} U^4 \biggr).  &   \cr
 &            &  (4.1.4) \cr
}$$

\noindent We use the short notation: $ \Delta E \equiv E_0 - \mu$.

\vskip 1cm

\noindent {\bf 4.2. Calculation of Tr[${\bf K}^5$]}   \par

For $n=5$, the expression of $Tr[{\bf K}^5]$ obtained from the
procedure {\tt perm}, where each term is multiplied by the respective
constant, is:   \par

\vskip -0.7cm

$$ \eqalignno{
Tr[{\bf K}^5]  &=
 5 < {\cal U}, {\cal E}_0, {\cal E}_0, {\cal E}_0, {\cal E}_0> +
5 < {\cal U}, {\cal E}_0, {\cal U}, {\cal E}_0, {\cal E}_0> +
5 < {\cal U}, {\cal U}, {\cal E}_0, {\cal E}_0, {\cal E}_0> + &  \cr
 & + 5 < {\cal U}, {\cal U}, {\cal E}_0, {\cal U}, {\cal E}_0> +
10 < {\cal U}, {\cal U}, {\cal T}^-, {\cal T}^+, {\cal E}_0> +
5 < {\cal U}, {\cal U} , {\cal U}, {\cal E}_0, {\cal E}_0> +   &  \cr
 &  + 5  < {\cal U}, {\cal U}, {\cal U}, {\cal U}, {\cal E}_0> +
< {\cal U}, {\cal U}, {\cal U}, {\cal U}, {\cal U}> +
< {\cal E}_0, {\cal E}_0, {\cal E}_0, {\cal E}_0, {\cal E}_0> +   &  \cr
  &  +10 < {\cal T}^-, {\cal T}^+, {\cal T}^-, {\cal T}^+, {\cal E}_0> +
10 < {\cal T}^-, {\cal T}^+, {\cal T}^-, {\cal T}^+, {\cal U}> +
10 < {\cal T}^-, {\cal U}, {\cal U}, {\cal T}^+, {\cal E}_0> +  &  \cr
& + 10 <{\cal T}^-, {\cal T}^+, {\cal E}_0, {\cal U}, {\cal E}_0> +
10 < {\cal T}^-, {\cal T}^+, {\cal E}_0, {\cal E}_0, {\cal E}_0>+
10 < {\cal T}^-, {\cal T}^+, {\cal U}, {\cal E}_0, {\cal E}_0> +  &  \cr
& + 10 < {\cal T}^-, {\cal T}^+, {\cal U}, {\cal U}, {\cal E}_0> +
10 < {\cal T}^-, {\cal T}^+, {\cal U}, {\cal U}, {\cal U}> +
10 < {\cal T}^-, {\cal E}_0, {\cal T}^+, {\cal U}, {\cal E}_0> +   &  \cr
&  + 10 < {\cal T}^-, {\cal E}_0, {\cal T}^+, {\cal E}_0, {\cal E}_0> +
10 < {\cal T}^-, {\cal U}, {\cal T}^+, {\cal E}_0, {\cal E}_0> +
10 < {\cal T}^-, {\cal U}, {\cal T}^+, {\cal U}, {\cal E}_0> +  &  \cr
&+ 10 < {\cal T}^-, {\cal U}, {\cal T}^+, {\cal U}, {\cal U}> +
10 < {\cal T}^+, {\cal T}^-, {\cal T}^-, {\cal T}^+, {\cal E}_0> +
10 < {\cal T}^+, {\cal T}^-, {\cal T}^-, {\cal T}^+, {\cal U}> +  &  \cr
& + 10 < {\cal U}, {\cal T}^-, {\cal U}, {\cal T}^+, {\cal E}_0> +
10 < {\cal U}, {\cal T}^-, {\cal E}_0, {\cal T}^+, {\cal E}_0> +
10 < {\cal U}, {\cal T}^-, {\cal T}^+, {\cal E}_0, {\cal E}_0> +  &  \cr
& + 10 < {\cal U}, {\cal T}^-, {\cal T}^+, {\cal U}, {\cal E}_0> +
10 < {\cal T}^-, {\cal T}^-, {\cal T}^+, {\cal T}^+, {\cal E}_0> +
10 < {\cal T}^-, {\cal T}^-,{\cal T}^+, {\cal T}^+, {\cal U}>, &  \cr
  &     &  (4.2.1) \cr
}$$

\vskip -0.5cm

\noindent where we are using the convention (4.1.2a) to write down
each term.  \par

In $n=5$ we have seven new type  of integrals  that do not have an
equivalent one for $n<5$. We table those integrals in section C.2
of Appendix C.     \par

After a long but convergent calculation, we get    \par

\vskip -0.7cm

$$ \eqalignno{
Tr[{\bf K}^5]  &= N^5 \biggl( {1 \over 1024} U^5 +
{5\over 256} U^4 \Delta E  + {5\over 4} U \Delta E^4 +
{5\over 32} U^3 \Delta E^2 + \Delta E^5 +
 {5\over 8} U^2 \Delta E^3 \biggr)
+   &   \cr
& + N^4 \biggl( {105\over 16} U^2 \Delta E^3  +
 {15\over 16} U^2 \Delta E \lambda_B^2 + {5\over 32} U^3 t^2 +
10 t^2 \Delta E^3 + 5 \Delta E^3 \lambda_B^2 + 5 \Delta E^5 +
{35 \over 4} U \Delta E^4 +   &   \cr
& + {15 \over 4} U \Delta E^2 \lambda_B^2 + 
{15 \over 2} U t^2 \Delta E^2 +
{15 \over 512} U^5 + {5 \over 64} U^3 \lambda_B^2 +
{155\over 64} U^3 \Delta E^2 + {55\over 128} U^4 \Delta E +   &   \cr
& + {15\over 8} U^2 t^2 \Delta E \biggr) + 
N^3 \biggl( 15 t^2 \Delta E^3 +
15 t^2 \Delta E \lambda_B^2 + {195\over 16} U \Delta E^4 +
{45\over 8} U \Delta E ^2 \lambda_B^2 +   &  \cr
& + {15\over 16} U \lambda_B^4 + 
{15 \over 2} \Delta E^3 \lambda_B^2+
{15 \over 4} \Delta E^5 + {15 \over 4} \Delta E \lambda_B^4 +
{15\over 4} U t^4 + {225 \over 16} U^2 \Delta E^3 +
{45\over 16} U^2 \Delta E \lambda_B^2 +   &  \cr
& + {15\over 4} U t^2 \lambda_B^2 + {75\over 4} U t^2 \Delta E^2 +
15 t^4 \Delta E + {495\over 256} U^4 \Delta E +
 {15\over 32} U^3 \lambda_B^2 + {15 \over 2} U^3 \Delta E^2 + &   \cr
&+ {75\over 8} U^2 t^2 \Delta E + {195\over 1024} U^5 +
{45\over 32} U^3 t^2 \biggr) + 
N^2 \biggl( -{25\over 4} U^2 t^2 \Delta E
+ {5\over 4} U^2 \Delta E^3  - &   \cr
& - {15\over 4} U^2 \Delta E \lambda_B^2 + {75\over 512} U^5 -
{15\over 16} U \Delta E^4 - {75\over8} U \Delta E^2 \lambda_B^2 -
{35 \over 16} U \lambda_B^4 - {15\over 2} t^4 \Delta E -   &  \cr
& - 15 U t^2 \Delta E^2 - {15 \over 2} U t^2 \lambda_B^2 +
{115\over 128} U^4 \Delta E - {15\over 2} \Delta E^3 \lambda_B^2 -
{5\over 4} \Delta E^5 - {5\over 4} \Delta E \lambda_B^4-
15 t^2 \Delta E^3 -  &    \cr
&- 15 t^2 \Delta E \lambda_B^2 - {45\over 64} U^3 \lambda_B^2 +
{125\over 64} U^3 \Delta E^2 - {15\over 8} U t^4 - {5\over 8} U^3 t^2
\biggr) +   &   \cr
& + N \biggl( - {5\over 2} U^2 \Delta E^3 -
 {15\over 2} U t^2 \Delta E^2 +
{15\over 2} U t^2 \lambda_B^2 - {15 \over 2} U^2 t^2 \Delta E -
{25\over 32}  U^4 \Delta E - {65\over 32} U^3 \Delta E^2 +  &  \cr
& + {5 \over 32} U^3 \lambda_B^2 - {15\over 128} U^5 - 
{15\over 8} U^3 t^2
- {5\over 4} U \Delta E^4  + {5\over 4} U \lambda_B^4 \biggr). &  \cr
&        &  (4.2.2) \cr
}$$

\noindent We continue to use the notation: $ \Delta E \equiv E_0 - \mu$.

 Certainly, the most subtle part in calculating
expression (4.2.2) comes from the product of Grassmann integrals
for different $\sigma$--sectors when we have Grassmann generators
at the same space indice. In this case, we have to suit
the conditions satisfied by both integrals and calculate the
contribution of the product to the sum over space indices.   \par

\vskip 1cm

\noindent {\bf 4.3. The $\beta$--Expansion of the Grand Potential
Up to Order $ \beta^4$ and     \hfill \break
\phantom{...} \hskip 0.4cm Physical Quantities}  \par

The relation between the grand potential  ${\cal W}(\beta; \mu)$
and the grand canonical partition function ${\cal Z}(\beta; \mu)$
is   \par

$$ {\cal W}(\beta;\mu) =- {1 \over \beta}
		 \ln {\cal Z}(\beta; \mu). \eqno(4.3.1) $$

The $\beta$--expansion of ${\cal Z}(\beta; \mu)$ up to order
$\beta^5$ is (see eq.(2.2)):  \par

\vskip -0.5cm

$$ {\cal Z}(\beta, \mu) \approx
 {\rm Tr}[ 1\!\hskip -1pt{\rm I} - \beta {\bf K}] +
{\beta^2 \over 2!} {\rm Tr}[ {\bf K}^2] -
 {\beta^3 \over 3!} {\rm Tr}[{\bf K}^3]  +
%
%
 {\beta^4 \over 4!} {\rm Tr}[ {\bf K}^4] -
{\beta^5 \over 5!} {\rm Tr}[{\bf K}^5].    \eqno (4.3.2) $$

\noindent The first term on the r.h.s. of (4.3.2) was calculated in
reference[11], the second and third terms were calculated in
reference [2] and in its last two terms we substitute the
results of eqs.(4.1.4) and (4.2.2).   \par

From eqs.(4.3.1) and (4.3.2), we get the grand potential up
to order $\beta^4$, that is,   \par

$$ \eqalignno{
{\cal W}(\beta;\mu) & = - N \Bigl\{ {2\over \beta} \ln 2 +
 \biggl( - {1\over 16} U t^2 \lambda_B^2 + {1\over 16} U^2 t^2 \Delta E
+ {1\over 1024} U^5 + {1\over 16} U t^2 \Delta E^2 +
{13 \over 768} U^3 \Delta E^2  -   &  \cr
& - {1 \over 768} U^3 \lambda_B^2 - {1\over 96} U \lambda_B^4 +
{1\over 96} U \Delta E^4 + {5\over 768} U^4 \Delta E +
 {1\over 48} U^2 \Delta E^3  + {1\over 64} U^3 t^2 \biggr) \beta^4-
   &     \cr
& -  \biggl( {1\over 8} t^2 U \Delta E + {1\over 16} U \Delta E \lambda_B^2
+ {1\over 96} \Delta E^4 + {1\over 16} t^4 + {1\over 96} \lambda_B^4
+ { 1\over 1024} U^4 + {1\over 48} U \Delta E^3
+   &   \cr
&  + {1\over 8} t^2 \Delta E^2 +
  {1 \over 192} U^3 \Delta E + {5 \over 96} t^2 U^2 +
{1\over 64} U^2 \lambda_B^2 + {1\over 64} U^2 \Delta E^2 +
{1 \over 16} \Delta E^2 \lambda_B^2  +  &    \cr
& + {1\over 8} t^2 \lambda_B^2 \biggr)\beta^3 +
 \biggl( -{U^3\over 64} + {1\over 16} U \lambda_B^2
- {1\over 16} \Delta E^2 U - {1\over 16} \Delta E U^2 \biggr) \beta^2
+ & \cr
 &  \hskip -0.5cm
+ \biggl( {1\over 4} \Delta E^2 + {1\over 4}
\lambda_B^2 + {1\over 4} \Delta E\; U + {t^2 \over 2} + {3\over 32}
U^2 \biggr) \beta - \big( \Delta E + {U\over 4} \bigr)  +
{\cal O}(\beta^5) \Bigr\} . & \cr
&    \cr
         &       &  (4.3.3) \cr
}$$

\noindent It is important to stress out that the coefficients of
the $\beta$--expansion of function ${\cal W}(\beta;\mu)$ are
exact for any set of constants: $E_0$, $t$, $U$, $\mu$ and $B$
for the unidimensional Hubbard model. From expression (4.3.3)
we can get the strong limit  approximation by taking
$U \gg t$, as well the atomic limit approximation when 
$U \ll t$. \par

From expression (4.3.3), we can derive any physical quantity for the
model at thermal equilibrium at high temperature.  As examples,
we consider the following quantities:  \par

\vskip 0.5cm

\noindent $i)$ specific heat at constant length and constant number
of fermions: $C_{L} (\beta)$. \par

$$ C_{L} (\beta) = - k \beta {\partial \over \partial \beta}\biggl[ \beta^2
{\partial {\cal W}(\beta;\mu) \over \partial \beta}\biggr],
 \eqno (4.3.4)  $$

\noindent where $k$ is the Boltzmann constant. From eq.(4.3.3), we
get,

\vskip -0.5cm

$$ \eqalignno{
C_{L} (\beta)& = Nk \biggl\{ \biggl( {5\over 256} U^5 -
 {5\over 24} U \lambda_B^4
+ {5 \over 24} U \Delta E^4 + {65\over 192} U^3 \Delta E^2 +
{5\over 12} U^2 \Delta E^3 - {5\over 192} U^3 \lambda_B^2 -  &  \cr
& - {5\over 4} U t^2 \lambda_B^2 + {5\over 4} U t^2 \Delta E^2 +
{5\over 4} U^2 t^2 \Delta E + {25 \over 192} U^4 \Delta E +
{5 \over 16} U^3 t^2 \biggr) \beta^5 + \biggl( -{3 \over 256} U^4 - & \cr
& - {3\over 2} t^2 \Delta E^2 - {1\over 8} \lambda_B^4 - {3 \over 4} t^4
- {1 \over 8} \Delta E^4 - {5\over 8} t^2 U^2 - {1\over 16} U^3 \Delta E
- {3\over 4} U \Delta E \lambda_B^2 -
 {3\over 4} \Delta E^2 \lambda_B^2  -  &  \cr
& - {3\over 2} t^2 \lambda_B^2 - {3\over 2} t^2 U \Delta E -
{3\over 16} U^2 \lambda_B^2 - {3\over 16} U^2 \Delta E^2 -
{1 \over 4} U \Delta E^3 \biggr)  \beta^4 +   &   \cr
& + \biggl( - {3\over 8} U^2 \Delta E - {3\over 32} U^3 +
{3\over 8} U \lambda_B^2 - {3\over 8} U \Delta E^2 \biggr) \beta^3 +  &\cr
& + \biggl( {1\over 2} U \Delta E + t^2 + {3\over 16} U^2 +
{1\over 2} \lambda_B^2 + {1\over 2} \Delta E^2 \biggr) \beta^2
+ {\cal O}(\beta^6) \biggr\}.
&  (4.3.4a)
}$$

\vskip 0.5cm

\noindent $ii)$ average energy per site: $<h> (\beta)$. \par

The simplest way to derive the average energy per site from the
grand potential, is to scale the constants: $(E_0, t, U,
\lambda_B) \rightarrow (\alpha E_0, \alpha t, \alpha U,\alpha
\lambda_B)$ and substitute in eq.(3.2a) to obtain
${\cal W}(\beta; \mu; \alpha)$. From the scaled grand potential,
we have \par

$$ <h> (\beta) =  {1\over  N} {\partial {\cal W}(\beta;\mu; \alpha)
\over \partial \alpha} \Big|_{\alpha=1}. \eqno(4.3.5) $$

\noindent  From eq.(4.3.3), we get that \par

\vskip -0.5cm

$$ \eqalignno{
<h> (\beta) & = E_0 + {U\over 4} +
 \biggl( - t^2 -{3\over 16} U^2 -{1\over 2} E_0 U + {1\over
4}U \mu - {1\over 2} E_0^2 + {1\over 2} E_0 \mu - {1\over 2}
\lambda_B^2 \biggr) \beta + & \cr
 & \hskip -0.5cm +
\biggl(  {3\over 16} E_0 U^2 -
{1\over 8} U^2 \mu - {3\over 16} U \lambda_B^2 + {3\over 16} U E_0^2 -
{1\over 4} U E_0 \mu +
{1\over 16} U \mu^2 + {3\over 64} U^3  \biggr) \beta^2 + &  \cr
& + \biggl( - {3 \over 16} U E_0^2 \mu + {1\over 256} U^4 + {1\over 4} t^4+
{1\over 24} \lambda_B^4 + {1\over 16} U^2 \lambda_B^2 +
{5 \over 24}t^2 U^2 + {1\over 2} t^2 \lambda_B^2 +  &  \cr
& + {1 \over 16} U^2 E_0^2 + {1\over 32} U^2 \mu^2 + {1\over 12} U E_0^3
- {1\over 48} U \mu^3 + {1\over 48} U^3 E_0 - {1\over 64} U^3 \mu +
{1 \over 24} E_0^4 -  &  \cr
& - {3 \over 32} U^2 E_0 \mu +  {1\over 8} U E_0 \mu^2 +
 {1\over 2} U t^2 E_0 - {3\over 8} U t^2 \mu - {1\over 8} E_0^3 \mu +
{1\over 8} E_0^2 \mu^2 -   &  \cr
& - {1\over 24} E_0 \mu^3 + {1\over 2} t^2 E_0^2 + {1\over 4} t^2 \mu^2 -
{3\over 4} t^2 E_0 \mu + {1\over 4} U E_0 \lambda_B^2 -
{3\over 8} \lambda_B^2 E_0 \mu - &  \cr
& -  {3\over 16} U \lambda_B^2 \mu + {1\over 4} \lambda_B^2 E_0^2 +
{1\over 8} \lambda_B^2 \mu^2 \biggr) \beta^3 +  &  \cr
& + \biggl( - {5\over 1024} U^5 + - {25\over 768} U^4 E_0 -
{5 \over 64} U^3 t^2 + {5\over 96} U \lambda_B^4 +
{5\over 192} U^4 \mu - {5 \over 48} U^2 E_0^3 +  &   \cr
& + {5 \over 768} U^3 \lambda_B^2
 - {5 \over 96} U E_0^4 - {1\over 96} U \mu^4 - {65 \over 768} U^3 E_0^2
+ {1\over 24} U^2 \mu^3 + {1\over 6} U E_0^3 \mu -   &   \cr
&  + {13\over 96} U^3 E_0 \mu
- {3\over 16} U E_0^2 \mu^2 + {1\over 4} U^2 E_0^2 \mu -
{3 \over 16} U^2 E_0 \mu^2 + {1 \over 12} U E_0 \mu^3 -
{5 \over 16} U^2 t^2 E_0 +   &  \cr
& + {5 \over 16} U t^2 \lambda_B^2 - {5\over 16} U t^2 E_0^2 -
{3\over 16} U t^2 \mu^2 + {1\over 4}t^2 U^2 \mu +{1\over 2} U t^2 E_0 \mu
- {13 \over 256} U^3 \mu^2 \biggr) \beta^4 +  &  \cr
& \hskip 2cm + {\cal O}(\beta^5).  & (4.3.5a)\cr }$$

\vskip 0.5cm

\noindent $iii)$ difference between average numbers of spin up and spin
down particles per site:   \hfill \break
$< n_\uparrow> - <n_\downarrow>$. \par

From the definition of the grand potential (eq.(4.3.1)), we have
that \par

$$ < n_\uparrow> (\beta) - <n_\downarrow> (\beta) = {1 \over  N}
{ \partial {\cal W}(\beta;\mu) \over \partial \lambda_B}.  \eqno(4.3.6)
$$

\noindent  Up to order $\beta^4$, we get from eq.(4.3.3) that, \par

\vskip -0.5cm

$$ \eqalignno{
 < n_\uparrow> (\beta) - <n_\downarrow> (\beta) &= -{\lambda_B \over8}
\biggl\{ - \bigl( U \beta + 4 \bigr) \beta + \biggl( \Delta E^2 +
U \Delta E + {1\over 3} \lambda_B^2 + {1\over 4} U^2
+ 2 t^2   \biggr) \beta^3 +
&    \cr
&  \hskip -1.5cm
 + \biggl( {1\over 3} U \lambda_B^2 + U t^2 + {1\over 48} U^3 \biggr)
\beta^4 +  {\cal O}(\beta^5) \biggr\}. & (4.3.6a)   \cr
} $$

\vskip 0.5cm

\noindent $iv)$ average of the square of the magnetization per site:
 $< m_z^2> (\beta) $.\par

\vskip -0.5cm

$$ \eqalignno{ < m_z^2> (\beta) &= {\lambda_B \over B^2} <({\bf n}_{i
\uparrow} - {\bf n}_{i \downarrow})^2 > & \cr
%
%
 & =
- \big( {1\over 2} g \mu_B \big)^2 {1\over  N} \Big[ {\partial
{\cal W}(\beta;\mu) \over \partial \mu} + 2 {\partial {\cal
W}(\beta;\mu) \over \partial U} \Big], & (4.3.7) \cr }$$

\noindent where $B$ is the external magnetic field. From eq.(4.3.3),
we obtain that \par

\vskip -0.5cm

$$\eqalignno{
< m_z^2> (\beta) &= {1\over 4} g^2 \mu_B^2 \biggl\{
{1\over 2} + {1\over 8} U \beta +
 \bigl( -{1\over 8} U E_0 + {1\over 8} \mu U - {1\over 32} U^2 + {1\over 8}
\lambda_B^2 - {1\over 8} E_0^2 +   &    \cr
& + {1\over 4} E_0 \mu - {1\over 8} \mu^2 \bigr) \beta^2  -
 \biggl( { 1\over 12} U t^2 + {1 \over 384} U^3 \biggr) \beta^3 +  &  \cr
& + \biggl( {5 \over 1536} U^4 + {15\over 384} U^2 E_0^2 +
{15\over 384} U^2 \mu^2 + {1\over 24} U E_0^3 - {1\over 24} U \mu^3 +
&     \cr
& + {7 \over 384} U^3 E_0 - {7 \over 384} U^3 \mu + {1\over 32} t^2 U^2
- {3\over 384} U^2 \lambda_B^2 - {1\over 8} t^2 \lambda_B^2 -
{1 \over 48} \lambda_B^4 - {1\over 8} U E_0^2 \mu +   &  \cr
& + {1\over 8} U E_0 \mu^2 + {1\over 8} U t^2 E_0 - {1\over 8} U t^2 \mu +
{1\over 48} E^4 + {1\over 48} \mu^4 - {1\over 12}E_0^3\mu +
{1\over 8} E_0^2 \mu^2 -  &   \cr
 &- {1 \over 12} E_0 \mu^3 + {1\over 8} t^2 E_0^2 + {1\over 8}t^2 \mu^2
- {1\over 4} t^2 E_0 \mu - {15 \over 192} U^2 E_0 \mu \biggr) \beta^4
 + {\cal O}(\beta^5) \biggr\} , & (4.3.7a) \cr }$$

\noindent where {\it g} is the Land\'e's factor and $\mu_B$ is the
Bohr's magneton. \par

\vskip 0.5cm

\noindent $v)$  magnetic susceptibility: $\chi (\beta)$.  \par

$$ \chi (\beta) = - \big( {1\over 2} g \mu_B\bigr)^2  \;{1 \over N}
{\partial ^2 {\cal W}(\beta;\mu)  \over \partial \lambda_B^2 }.
\eqno (4.3.8)   $$

\noindent From eq. (4.3.3), we obtain that  \par

$$\eqalignno{
\chi (\beta) &= - \bigl( {1\over 2} g \mu_B\bigr)^2 \;\biggl[
\biggl( {1\over 8} U t^2 + {1\over 8} U \lambda_B^2 + {1\over 384} U^3
\biggr)  \beta^4  +  \biggl( {1\over 4} t^2 + {1\over 8} \lambda_B^2 +
{1\over 8} U \Delta E +    &  \cr
& + {1\over 8} \Delta E^2 + {1\over 32} U^2
\biggr)  \beta^3 -  {1\over 8} U \beta^2 -
{1\over 2} \beta + {\cal O}(\beta^5)  \biggr].   &  (4.3.8a) \cr
}$$

\vskip 1cm

\noindent {\bf 5. Conclusions}

\bigskip

With the implementation of the factorization into sub--graphs of the
Grassmann multivariable integrals, we can certainly go beyond the
calculation of the term $\beta^4$ of the $\beta$--expansion of the
grand potential of the unidimensional Hubbrad model. Even though 
the physics for $U>0$ and $U<0$ are different, the results
of section 4.3 apply equally well for both cases. Recently,
dos Santos and Thomaz have applied the results of reference [2] to 
calculate the $\beta$--expansion of the grand canonical partition
function of the extended unidimensional Hubbard model up to
orde $\beta^3$ [16]. But the important
point is that the present approach opens the possibility to calculate
the first terms of the $\beta$--expansion of the grand canonical
partition function of the Hubbard model in two space dimensions, as
well as of unidimensional models with impurities. We believe that
improvements on the present approach will render a valuable tool for
tackling with such problems.

\vfill
\eject

%
%

\centerline {\bf Appendix A}  \par

\centerline{\bf Factorization of Grassmannian Sub--Graphs}  \par

\vskip 0.5cm
For calculating  the co--factors of  matrices
${\bf A}^{\sigma \sigma}$, $ \sigma= \uparrow, \downarrow$, it helps to
have the value of their determinant. For arbitrary $n$, the determinant
of these matrices  is equal to  \par

$$ \det {\bf A}^{\sigma \sigma} = \Bigl[\;\; \prod_{\nu =0}^{n-1}
\lambda_{\nu}^{(n)}\;\; \Bigr] ^N,   \eqno (A.1)
$$

\noindent  where $N$ is the number of space sites and $\lambda_{\nu}^{(n)}$
are the {\it N}--fold degenerated eigenvalue of
${\bf A}^{\sigma \sigma}$. From eq.(2.6e) we have that  \par

$$
\lambda_\nu^{(n)} = 1 - e^{{i\pi \over n} (2\nu +1)}.  \eqno(A.2)
$$

\noindent We should notice that
$ \lambda_{(n-1)-\nu}^{(n)} = {\lambda_\nu^{(n)}}^{*}$.   \par

We define:  \par

$$  P^{(n)} \equiv \prod_{\nu =0}^{n-1}  \lambda_{\nu}^{(n)}.   \eqno(A.3)
$$

For calculating $P^{(n)}$ we need to consider  the cases $n$ even  and
$n$ odd separately.  \par

For $n$ even, expression (A.3) can be rewritten as,   \par

$$  P^{(n)} = \prod_{ \nu =0}^{n-2 \over 2} \; \biggl[
2 - 2 \cos \bigl({\pi\over n} (2\nu +1) \bigr) \;\; \biggr] = 2,
	\eqno(A.4)  $$

\noindent where the last equality is already known [13].  \par

For $n$ odd, expression (A.3) can be rewritten as,  \par

$$ P^{(n)} = 2 \times 2^{n-1} \;\; \prod_{ \nu =0}^{n-1} \;\;
\sin \biggl( {\pi \over 2n}(2 \nu +1) \biggr).  \eqno(A.5)
$$

\noindent From reference [13], we have that [14]:   \par

$$ 2^{n-1} \;\; \prod_{ \nu =0}^{n-1} \;\;
\sin \biggl( {\pi \over 2n}(2 \nu +1) \biggr) = 1,  \eqno(A.6) $$

\noindent that substituted  in eq. (A.5) gives \par

$$ P^{(n)} = 2.   \eqno (A.7)   $$

From the results (A.4) and (A.7), for any $n$, we get that  \par

$$ \eqalignno{
 \det {\bf A}^{\sigma \sigma} & = \Bigl[\;\; \prod_{\nu =0}^{n-1}
\lambda_{\nu}^{(n)}\;\; \Bigr] ^N   &   \cr
& = 2^N.    & (A.8)  \cr
}   $$

\vskip 0.5cm

To present the factorization of the Grassmannian integrals, we consider
an example and use the graphic notation explained  in Appendix B.

Let us consider  the integral for fixed space indices $l_1$ and $l_3$,
that contributes to  \hfill\break
$< {\cal E}_0 (\uparrow), {\cal E}_1 (\uparrow),{\cal E}_2 (\uparrow)>$,
 \par

\vskip -0.7cm

$$  
 {\cal I} (l_1, l_1, l_3)   =
\int \prod_{I=1}^{4N}
\, d\eta_I d\bar{\eta}_I \;\; e^{\sum\limits_{I,J= 1}^{4N}
\bar{\eta}_I\; A_{I J}^{\uparrow\uparrow} \; \eta_J} 
\;\; \bar{\eta}_{l_1} \;\eta_{l_1}
\;\;\bar{\eta}_{N+l_1}\; \eta_{N+l_1} \;\; \bar{\eta}_{2N+l_3} \;
\eta_{2N+l_3}  $$


\hskip 3cm \epsfysize = 3cm

\vskip -2cm
$$\eqalignno
 {&     &    (A.9) \cr}  $$

\noindent where $l_1 \not= l_3$. Under the similarity transformation
(2.5), eq.(A.9) becomes  \par

\vskip -0.7cm

$$ \eqalignno{
{\cal I} (l_1, l_1, l_3) & = \int \prod_{I=1}^{4N}
d\bar{\eta}_I^\prime d{\eta}_I^\prime \;\;
\Bigl[  \sum_{ {\nu_1, \nu_2 =0} \atop {\tau_1,\tau_2=0}}^3 \;\;
q_{\nu_1 0} p_{0 \tau_1} \; q_{\nu_2 1} p_{1 \tau_2}
\bar{\eta}_{\nu_1 N + l_1}^\prime \eta_{\tau_1 N + l_1}^\prime
\bar{\eta}_{\nu_2 N + l_1}^\prime  \eta_{\tau_2 N + l_1}^\prime
\;\; \Bigr] \times &   \cr
%
%
&\hskip 2cm    \times \Bigl[ \sum_{\nu_3 = 0}^3 \; \;
q_{\nu_3 2} p_{2 \nu_3 }
\bar{\eta}_{\nu_3 N + l_3}^\prime  \eta_{\nu_3 N + l_3}^\prime \Bigr]
e^{\sum\limits_{I,J= 1}^{4N}
\bar{\eta}_I\; D_{I J} \; \eta_J}    &   \cr
%
%
& =
\Bigl[ 2^N \;  \sum_{ {\nu_1, \nu_2 =0} \atop {\tau_1,\tau_2=0}}^3 \;\;
{ q_{\nu_1 0} p_{0 \tau_1} \; q_{\nu_2 1} p_{1 \tau_2}    \over
\lambda_{\nu_1}^{(4)} \lambda_{\nu_2}^{(4)} } \Bigr]
\times
\Bigl[ \sum_{\nu_3 = 0}^3 \; \;
{q_{\nu_3 2} p_{2 \nu_3 } \over \lambda_{\nu_3}^{(4)}} \Bigr],
&   (A.10)  \cr
}$$

\noindent where to write the second equality on the r.h.s. of eq.(A.10),
we used the result (A.8).   \par

By explicitly writting down the expressions, we see that,  \par

\vskip -0.5cm

$$ \eqalignno{
& 2^N \;  \sum_{ {\nu_1, \nu_2 =0} \atop {\tau_1,\tau_2=0}}^3 \;\;
{ q_{\nu_1 0} p_{0 \tau_1} \; q_{\nu_2 1} p_{1 \tau_2}    \over
\lambda_{\nu_1}^{(4)} \lambda_{\nu_1}^{(4)} } =
\int \prod_{I=1}^{4N}
\, d\eta_I d\bar{\eta}_I \;\; e^{\sum\limits_{I,J= 1}^{4N}
\bar{\eta}_I\; A_{I J}^{\uparrow\uparrow} \; \eta_J}
\;\; \bar{\eta}_{l_1} \;\eta_{l_1}
\;\;\bar{\eta}_{N+l_1}\; \eta_{N+l_1} ,  &  \cr
      &      & (A.10a)\cr
%
%
\noalign{\hbox{and}}
%
%
 & \sum_{\nu_3 = 0}^3 \; \;
{q_{\nu_3 2} p_{2 \nu_3 } \over \lambda_{\nu_3}^{(4)}}  = {1 \over 2^N}\;
\int \prod_{I=1}^{4N}
\, d\eta_I d\bar{\eta}_I \;\; e^{\sum\limits_{I,J= 1}^{4N}
\bar{\eta}_I\; A_{I J}^{\uparrow\uparrow} \; \eta_J}
\;\;\bar{\eta}_{2N+l_3}\; \eta_{2N+l_3}.   & (A.10b)\cr
}$$

\noindent Using the graphic representation of Appendix B,
 we write result (A.10) as,  \par

\vskip 0.3cm

\epsfysize= 3cm

\vskip -2.5cm

$$  \eqalignno{  &     &  (A.11)\cr
       &       &   }  $$


\vskip -0.6cm

The factorization (A.11) comes directly from the fact that the matrix
{\bf D} is diagonal (see eq.(2.5a)) and  the result (2.7a). Once
the presence of Grassmann generators in the integrand of integrals
(2.8) correspond to cutting lines and columns of matrix {\bf D},then
only for cutts at the same space index and any $\nu$--indices  we
get co--factors of matrix {\bf D} that are non--zero. In summary,
the factorization of the type (A.11) always happens when two or
more space indices are different.   \par

In a similar way and by the reasons discussed before, it is simple to
show that ${\cal I}(l_1, l_2, l_3)$, where all the space indices
$l_i$, $i=1,2,3$, are distinct, is easily written as:  \par


\vskip 0.5cm

\epsfysize= 3cm

\vskip -2.5cm

$$   \eqno(A.12)  $$


\vfill
\eject


\centerline {\bf Appendix B}  \par

\centerline{\bf Graphic Notation of the Multivariable
     Grassmann Integrals}  \par

\vskip 0.5cm

To exemplify our graphic notation for the graphs that contribute to
eq. (2.3), for fixed value of $n$, we consider some terms of
$Tr[{\bf K}^4]$. This graphic notation is very helpful when we apply
the symmetries discussed in reference [2] to identify equivalent
terms in  $Tr[{\bf K}^n]$.  \par

We present the graphic notation through examples and its application
 to the identification of equivalent integrals.   \par

\vskip 0.5cm

\noindent {\bf 1)}   \par

$$ \hskip -4.5cm <{\cal E}(\uparrow,0), {\cal E}(\downarrow,1),
{\cal E}(\uparrow,2), {\cal E}(\uparrow,3)>  =  
E(\uparrow)^3 E(\downarrow) \times 
$$

\vskip -3cm

$$
$$

\vskip 1.5cm

\noindent {\bf 2)}   \par

$$ \hskip -4.5cm  <{\cal E}(\downarrow,0) {\cal E}(\uparrow,1)
{\cal E}(\uparrow,2) {\cal E}(\uparrow,3)>  =  
E(\uparrow)^3 E(\downarrow) \times 
$$

\vskip -3cm

$$
$$

\noindent The constants $E(\uparrow)$ and $E(\downarrow)$ are defined
just below eq.(3.4d). In order to show that terms (B.1) and (B.2) are
equal, we use the invariance of the integrals under a cyclic
translation in the temperature parameter $\nu$ in each sector $\sigma
= \uparrow$ and $\sigma = \downarrow$, separately. Therefore, \par

$$<{\cal E}(\uparrow,0) {\cal E}(\downarrow,1)
{\cal E}(\uparrow,2) {\cal E}(\uparrow,3)>  =
<{\cal E}(\downarrow,0), {\cal E}(\uparrow,1),
{\cal E}(\uparrow,2), {\cal E}(\uparrow,3)>  \eqno (B.3)  $$

\vfill
\eject

\noindent {\bf 3)}   \par

$$  \hskip -4.5cm
 < {\cal U}(0) {\cal T}^- (\uparrow, 1) {\cal T}^+(\uparrow, 2)
{\cal E} (\uparrow, 3) > = Ut^2 E(\uparrow) \times $$

\vskip -3cm

$$
$$

\vskip 0.2cm

\noindent Due to the presence of the term ${\cal U}(0)$, the integrals 
in the  two $\sigma$--sectors have one space index  $l$ in common.   \par

\vskip 1.5cm

\noindent {\bf 4)}   \par

$$\hskip -4.5cm
 < {\cal U}(0) {\cal T}^- (\downarrow, 1) {\cal T}^+(\downarrow, 2)
{\cal E} (\downarrow, 3) > = Ut^2 E(\downarrow) \times  $$

\vskip -3cm

$$
$$

\vskip 0.2cm

The terms (B.4) and (B.5) are equal, up to a multiplicative factor,
due to the fact that
${\bf A}^{\uparrow \uparrow} = {\bf A}^{\downarrow \downarrow}$. \par

\vskip 0.5cm

The graphic notation was used along all the calculations and
permitted us to considerably reduce the number of terms that
contribute to the expressions of
$Tr[{\bf K}^4]$ and $Tr[{\bf K}^5]$.

\vfill
\eject


\centerline {\bf Appendix C}  \par

\centerline{\bf Useful Multivariable Grassmann Integrals at
              $n=4$ and $n=5$}  \par

\vskip 0.5cm

\noindent {\bf C.1. Useful integrals for n=4}   \par

We need the result of twelve integrals only, to calculate the terms
that contribute to (4.1.3). In this Appendix, we present the value of
these integrals according to the conditions satisfied by the space
indices [15]. \par

\noindent 1) \par

\vskip -0.5cm

$$ {\cal I}_1^{(4)} (l) \equiv \int \prod_{I=1}^{4N} \, d\eta_I
d\bar{\eta}_I \;\; e^{\sum\limits_{I,J= 1}^{4N} \bar{\eta}_I\;
A_{I J}^{\uparrow \uparrow}
\; \eta_J} \;\; \bar{\eta}_l \;\eta_l \;\; = 2^{N-1}, \eqno(C.1.1) $$

\noindent for $ l= 1, 2, \cdots, N$. \par

\noindent 2) \par

\vskip -0.5cm

$$\eqalignno{ {\cal I}_2^{(4)} (l_1,l_2) &\equiv \int \prod_{I=1}^{4N} \,
d\eta_I d\bar{\eta}_I \;\; e^{\sum\limits_{I,J= 1}^{4N} \bar{\eta}_I\;
A_{I J}^{\uparrow \uparrow}
 \; \eta_J} \;\; \bar{\eta}_{l_1}\; \eta_{l_1}
\;\;\bar{\eta}_{N+l_2} \;\eta_{N+l_2} & \cr
 & & \cr
%
%
 & =\cases{ 2^{N-1}, &
$ l_1=l_2, \hskip 0.5cm l_1= 1, 2, \cdots, N$ \cr 2^{N-2}, & $ l_2\not=
l_1,\hskip 0.5cm l_1, l_2=1, 2, \cdots, N.$\cr } & (C.1.2) \cr }$$

\noindent 3) \par

\vskip -0.5cm

$$ \eqalignno{ {\cal I}_3^{(4)} (l_1,l_2) &\equiv \int \prod_{I=1}^{4N} \,
d\eta_I d\bar{\eta}_I \;\; e^{\sum\limits_{I,J= 1}^{4N} \bar{\eta}_I\;
A_{I J}^{\uparrow \uparrow}
 \; \eta_J} \;\; \bar{\eta}_{l_1} \;\eta_{l_1+1}
\;\;\bar{\eta}_{N+l_2}\; \eta_{N+l_2 -1} & \cr
 & & \cr
%
%
& = 2^{N-2},   l_2= l_1 +1, \hskip 0.5cm l_1= 1, 2, \cdots, N.
          & (C.1.3) \cr }$$

\noindent 4) \par

\vskip -0.5cm

$$ \eqalignno{ {\cal I}_4^{(4)} (l_1, l_2, l_3) & \equiv \int
\prod_{I=1}^{4N} \, d\eta_I d\bar{\eta}_I \;\; e^{\sum\limits_{I,J=
1}^{4N} \bar{\eta}_I\; A_{I J}^{\uparrow \uparrow}
 \; \eta_J} \;\; \bar{\eta}_{l_1}
\;\eta_{l_1} \;\;\bar{\eta}_{N+l_2} \; \eta_{N+l_2+1}\;\;
\bar{\eta}_{2N+l_3} \; \eta_{2N+l_3 -1} & \cr
 & & \cr
%
%
 & = \cases{ 2^{N-3}, &  $l_3=l_2 +1$\cr
	      2^{N-2}, & $l_1=l_3= l_2 -1$. \cr}
& (C.1.4) \cr }$$

\vfill
\eject

\noindent 5) \par

\vskip -0.5cm

$$\eqalignno{ {\cal I}_5^{(4)} (l_1, l_2, l_3) &\equiv \int \prod_{I=1}^{4N}
\, d\eta_I d\bar{\eta}_I \;\; e^{\sum\limits_{I,J= 1}^{4N}
\bar{\eta}_I\; A_{I J}^{\uparrow \uparrow}
 \; \eta_J} \;\; \bar{\eta}_{l_1} \;\eta_{l_1}
\;\;\bar{\eta}_{N+l_2}\; \eta_{N+l_2} \;\; \bar{\eta}_{2N+l_3} \;
\eta_{2N+l_3} & \cr
 & & \cr
%
%
 & =\cases{2^{N-3}, &  $l_1\not= l_2 \not= l_3$ \cr
     2^{N-2}, & $l_1= l_2 $, or, $l_1=l_3$, or, $l_2=l_3 $ \cr
              2^{N-1}, & $ l_1 = l_2 = l_3$. \cr }
& (C.1.5) \cr }$$

\noindent 6)      \par

\vskip -0.5cm

$$\eqalignno{ {\cal I}_6^{(4)} (l_1, l_2, l_3) &\equiv \int \prod_{I=1}^{4N}
\, d\eta_I d\bar{\eta}_I \;\; e^{\sum\limits_{I,J= 1}^{4N}
\bar{\eta}_I\; A_{I J}^{\uparrow \uparrow}
 \; \eta_J} \;\; \bar{\eta}_{l_1} \;\eta_{l_1 +1}
\;\;\bar{\eta}_{N+l_2}\; \eta_{N+l_2} \;\; \bar{\eta}_{2N+l_3} \;
\eta_{2N+l_3 -1} & \cr
 & & \cr
%
%
 & =\cases{2^{N-3}, & $l_3 = l_1 +1$   \cr
           2^{N-2}, & $l_2 = l_1 $ and  $l_3 = l_1 +1$. \cr  }
& (C.1.6) \cr }$$

\noindent 7)     \par

\vskip -0.5cm

$$\eqalignno{ {\cal I}_7^{(4)} (l_1, l_2, l_3) &\equiv \int \prod_{I=1}^{4N}
\, d\eta_I d\bar{\eta}_I \;\; e^{\sum\limits_{I,J= 1}^{4N}
\bar{\eta}_I\; A_{I J}^{\uparrow \uparrow}
 \; \eta_J} \;\; \bar{\eta}_{l_1} \;\eta_{l_1}
\;\;\bar{\eta}_{N+l_2}\; \eta_{N+l_2 +1} \;\; \bar{\eta}_{2N+l_3} \;
\eta_{2N+l_3 -1} & \cr
 & & \cr
%
%
 & =\cases{2^{N-3}, & $l_2 = l_3 -1$   \cr
           2^{N-2}, & $l_1 = l_3 = l_2 +1$. \cr  }
& (C.1.7) \cr }$$

\noindent 8)         \par

\vskip -0.5cm

$$\eqalignno{ {\cal I}_8^{(4)} (l_1, l_2, l_3, l_4)
            &\equiv \int \prod_{I=1}^{4N}
\, d\eta_I d\bar{\eta}_I \;\; e^{\sum\limits_{I,J= 1}^{4N}
\bar{\eta}_I\; A_{I J}^{\uparrow \uparrow}
 \; \eta_J} \;\; \bar{\eta}_{l_1} \;\eta_{l_1}
\;\;\bar{\eta}_{N+l_2}\; \eta_{N+l_2} \;\; \bar{\eta}_{2N+l_3} \;
\eta_{2N+l_3}  \;\;\bar{\eta}_{3N+l_4}\; \eta_{3N+l_4}  & \cr
 & & \cr
%
%
 & =\cases{ 2^{N-4}, & $l_1\not= l_2 \not= l_3 \not= l_4$ \cr
2^{N-3}, & $l_1 = l_2$, or,   $\cdots,{\rm or}, l_3 = l_4$ \cr
2^{N-2}, & $l_1= l_2 $ and  $ l_3=l_4$, or,   all permutations  2 by 2  \cr
   &$l_1=l_2=l_3$, or, all permutations with 3 equal
                                   space indices\cr
2^{N-1}, & $ l_1 = l_2 = l_3 = l_4$. \cr }
         & (C.1.8) \cr }$$

\vfill
\eject

\noindent 9)       \par

\vskip -0.5cm

$$\eqalignno{ {\cal I}_9^{(4)} (l_1, l_2, l_3, l_4)
&\equiv \int \prod_{I=1}^{4N}
\, d\eta_I d\bar{\eta}_I \;\; e^{\sum\limits_{I,J= 1}^{4N}
\bar{\eta}_I\; A_{I J}^{\uparrow \uparrow}
 \; \eta_J} \;\; \bar{\eta}_{l_1} \;\eta_{l_1+1}
\;\;\bar{\eta}_{N+l_2}\; \eta_{N+l_2-1} \;\; \bar{\eta}_{2N+l_3} \;
\eta_{2N+l_3+1}  \times  &  \cr
&  \hskip 6cm \times {\eta}_{3N+l_4}\; \eta_{3N+l_4 -1}  & \cr
 & & \cr
%
%
 &  \hskip -0.5cm
   = \cases{ 2^{N-4}, & $l_2 = l_1+1$ and $l_4= l_3+1$, or,
                        $l_2= l_3 +1$ and $l_4= l_1+1$ \cr
2^{N-2}, & $l_2= l_3+1 $ and  $ l_1=l_3$ and $ l_4=l_3+1$.   \cr}
         & (C.1.9) \cr }$$

\noindent 10)       \par

\vskip -0.5cm

$$\eqalignno{ {\cal I}_{10}^{(4)} (l_1, l_2, l_3, l_4)
         &\equiv \int \prod_{I=1}^{4N}
\, d\eta_I d\bar{\eta}_I \;\; e^{\sum\limits_{I,J= 1}^{4N}
\bar{\eta}_I\; A_{I J}^{\uparrow \uparrow}
 \; \eta_J} \;\; \bar{\eta}_{l_1} \;\eta_{l_1-1}
\;\;\bar{\eta}_{N+l_2}\; \eta_{N+l_2-1} \;\; \bar{\eta}_{2N+l_3} \;
\eta_{2N+l_3+1}  \times  &  \cr
&  \hskip 6cm \times {\eta}_{3N+l_4}\; \eta_{3N+l_4 +1}  & \cr
 & & \cr
%
%
 & =\cases{ 2^{N-4}, & $l_2 = l_3+1$ and $l_1= l_4 + 1$, or,
                        $l_4= l_2 - 1$ and $l_1= l_3+1$ \cr
2^{N-3}, & $l_2= l_3+1 $ and  $ l_1=l_3 +2$ and $ l_4=l_3+1$   \cr
          & $l_2 = l_3 +1$ and $l_1 = l_3$ and $ l_4 = l_3 -1$.  \cr}
         & (C.1.10) \cr }$$

\noindent 11)      \par

\vskip -0.5cm

$$\eqalignno{ {\cal I}_{11}^{(4)} (l_1, l_2, l_3, l_4)
 & \equiv \int \prod_{I=1}^{4N}
\, d\eta_I d\bar{\eta}_I \;\; e^{\sum\limits_{I,J= 1}^{4N}
\bar{\eta}_I\; A_{I J}^{\uparrow \uparrow}
 \; \eta_J} \;\; \bar{\eta}_{l_1} \;\eta_{l_1}
\;\;\bar{\eta}_{N+l_2}\; \eta_{N+l_2} \;\; \bar{\eta}_{2N+l_3} \;
\eta_{2N+l_3 +1} \times   &  \cr
&    \hskip 5cm \times  \;\;\bar{\eta}_{3N+l_4}\; \eta_{3N+l_4 -1}  & \cr
 & & \cr
%
%
 & \hskip -0.5cm
     =\cases{ 2^{N-4}, & $ l_4 = l_3 +1$ \cr
2^{N-3}, & $l_2 = l_3+1$ and $l_4 = l_3 +1$, or,
           $l_1 =l_2$  and $ l_4= l_3 +1$,     \cr
         & or,  $l_1= l_3 +1$ and $l_4 = l_3 +1$   \cr
2^{N-2}, & $l_1= l_3 + 1 $ and  $ l_2=l_3 +1$ and $ l_4= l_3 +1$.   \cr }
  & (C.1.11) \cr }$$

\noindent 12)     \par

\vskip -0.5cm

$$\eqalignno{ {\cal I}_{12}^{(4)}  (l_1, l_2, l_3, l_4)
       &\equiv \int \prod_{I=1}^{4N}
\, d\eta_I d\bar{\eta}_I \;\; e^{\sum\limits_{I,J= 1}^{4N}
\bar{\eta}_I\; A_{I J}^{\uparrow \uparrow}
 \; \eta_J} \;\; \bar{\eta}_{l_1} \;\eta_{l_1}
\;\;\bar{\eta}_{N+l_2}\; \eta_{N+l_2 +1} \;\; \bar{\eta}_{2N+l_3} \;
\eta_{2N+l_3}  \times &  \cr
& \hskip 5cm  \times \bar{\eta}_{3N+l_4}\; \eta_{3N+l_4 -1}  & \cr
 & & \cr
%
%
 &  \hskip -1.5cm
 =\cases{ 2^{N-4}, & $ l_4 = l_2 +1$ \cr
2^{N-3}, & $l_1 = l_3$ and $l_4 = l_2 +1$, or,
           $l_1 =l_2 +1$  and $ l_4= l_2 +1$,   \cr
         & or,  $l_2= l_3 $ and $l_4 = l_2 +1$   \cr
2^{N-2}, & $l_1= l_2 + 1 $ and  $ l_3=l_2$ and $ l_4= l_2 +1$.   \cr}
  & (C.1.12) \cr }$$

\vfill
\eject

\noindent {\bf C.2. Useful integrals for n=5}   \par

We present here the seven integrals that have no equivalent ones for
$n<5$; i.e., integrals that cannot be factorized into any of the
integrals for $n<5$ for all the conditions satisfied by the apace
indices.  In some graphs, we do not have generators in the integrand
of integrals of type (2.7) at a given
value of $\nu$, as we can see for example in the graphs
presented in Appendix B. Those rings in the integrals of type (2.7)
 that have no associated Grassmann  generators in the integrand, we
call {\it empty rings}. For example,
in (B.1) we have one empty ring at $\sigma=\uparrow$ ($\nu =1$),
  and three empty rings at $\sigma=\downarrow$ ($\nu= 0,1$, and $3$).
The integrals for $n=5$ with  empty rings give the same results
 to the equivalent integrals for $ n=4$.
We have not demonstrated this property in general form for any $n$,
but we have detected it by evaluating these integrals
through the procedure {\tt gint}.   \par

The seven integrals for $n=5$ and the conditions satisfied by the
space indices [15] are:

\noindent 1)    \par

\vskip -0.5cm

$$\eqalignno{ {\cal G}_1^{(5)} (l_1, l_2, l_3, l_4, l_5)
       & \equiv \int \prod_{I=1}^{5N}
\, d\eta_I d\bar{\eta}_I \;\; e^{\sum\limits_{I,J= 1}^{5N}
\bar{\eta}_I\; A_{I J}^{\uparrow \uparrow}
 \; \eta_J} \;\; \bar{\eta}_{l_1} \;\eta_{l_1}
\;\;\bar{\eta}_{N+l_2}\; \eta_{N+l_2} \;\; \bar{\eta}_{2N+l_3} \;
\eta_{2N+l_3}   \times  & \cr
& \hskip 3cm
 \times \bar{\eta}_{3N+l_4}\; \eta_{3N+l_4}
 \;\;\bar{\eta}_{4N+l_5}\; \eta_{4N+l_5} & \cr
 & & \cr
%
%
 &  \hskip -1cm
     =\cases{ 2^{N-5}, & $l_1 \not= l_2\not= l_3 \not= l_4 \not= l_5$  \cr
2^{N-4}, & $l_1= l_2$ , or, all permutations with 2 equal space indices\cr
2^{N-3}, & $l_1 = l_2= l_3$, or,  all permutations with 3 equal
			space indices  \cr
         & $l_1 = l_2$ and $ l_3 = l_4$,   \cr
         &    or, all permutations  2 by 2 \cr
2^{N-2}, & $l_1= l_2= l_3= l_5 $, or,   \cr
         &  all permutations with 4 equal space indices  \cr
         & $l_1= l_2= l_3$ and $l_4=l_5$, or, \cr
         & all permutations with 2 or 3  equal space indices   \cr
2^{N-1}, & $ l_1 = l_2 = l_3 = l_4 = l_5$. \cr }
         & (C.2.1) \cr }$$

\vskip 0.3cm

\noindent 2)   \par

\vskip -0.5cm

$$\eqalignno{ {\cal G}_2^{(5)} (l_1, l_2, l_3, l_4, l_5)
          & \equiv \int \prod_{I=1}^{5N}
\, d\eta_I d\bar{\eta}_I \;\; e^{\sum\limits_{I,J= 1}^{5N}
\bar{\eta}_I\; A_{I J}^{\uparrow \uparrow}
 \; \eta_J} \;\; \bar{\eta}_{l_1} \;\eta_{l_1}
\;\;\bar{\eta}_{N+l_2}\; \eta_{N+l_2+1} \;\; \bar{\eta}_{2N+l_3} \;
\eta_{2N+l_3}   \times   & \cr
& \hskip 3cm
 \times \bar{\eta}_{3N+l_4}\; \eta_{3N+l_4 -1}
 \;\;\bar{\eta}_{4N+l_5}\; \eta_{4N+l_5} & \cr
 & & \cr
%
%
 & =\cases{ 2^{N-5}, & $l_4 = l_2 + 1 $ \cr
2^{N-4}, & $l_4= l_2 + 1$ and two other space indices are equal\cr
2^{N-3}, & $l_4 = l_2+1$ and $l_1 = l_4$ and $l_3 = l_5$, \cr
         & or, all  permutations 2 by 2\cr
         & $l_4 = l_2+1$ and $ l_1 = l_4= l_5$, or,  \cr
         &  all permutations with 3 equal space indices  \cr
2^{N-2}, & $l_2 = l_5 -1= l_3$ and $l_1= l_5=l_4$.  \cr }
         & (C.2.2) \cr }$$

\vskip 0.3cm


\noindent 3)     \par

\vskip -0.5cm

$$\eqalignno{ {\cal G}_3^{(5)} (l_1, l_2, l_3, l_4, l_5)
          & \equiv \int \prod_{I=1}^{5N}
\, d\eta_I d\bar{\eta}_I \;\; e^{\sum\limits_{I,J= 1}^{5N}
\bar{\eta}_I\; A_{I J}^{\uparrow \uparrow}
 \; \eta_J} \;\; \bar{\eta}_{l_1} \;\eta_{l_1}
\;\;\bar{\eta}_{N+l_2}\; \eta_{N+l_2+1} \;\; \bar{\eta}_{2N+l_3} \;
\eta_{2N+l_3 -1}  \times & \cr
& \hskip 3cm
 \times \bar{\eta}_{3N+l_4}\; \eta_{3N+l_4}
 \;\;\bar{\eta}_{4N+l_5}\; \eta_{4N+l_5} & \cr
 & & \cr
%
%
 & =\cases{ 2^{N-5}, & $l_3 = l_2 + 1 $ \cr
2^{N-4}, & $l_3= l_2 + 1$ and two other space indices are equal\cr
2^{N-3}, & $l_3 = l_2+1$ and $l_1 = l_4$ and $l_3 = l_5$, or,  \cr
         & all permutations 2 by 2   \cr
         & $l_3 = l_2+1$ and $ l_1 = l_3= l_4$, or,   \cr
          & all permutations with 3 equal space indices \cr
2^{N-2}, & $l_3 = l_2 + 1$ and $l_1= l_3=l_4 = l_5$. \cr }
         & (C.2.3) \cr }$$

\vskip 0.3cm

\noindent 4)    \par

\vskip -0.5cm

$$\eqalignno{ {\cal G}_4^{(5)} (l_1, l_2, l_3, l_4, l_5)
         & \equiv \int \prod_{I=1}^{5N}
\, d\eta_I d\bar{\eta}_I \;\; e^{\sum\limits_{I,J= 1}^{5N}
\bar{\eta}_I\; A_{I J}^{\uparrow \uparrow}
 \; \eta_J} \;\; \bar{\eta}_{l_1} \;\eta_{l_1 -1}
\;\;\bar{\eta}_{N+l_2}\; \eta_{N+l_2+1} \;\; \bar{\eta}_{2N+l_3} \;
\eta_{2N+l_3 +1}  \times  & \cr
& \hskip 3cm
 \times \bar{\eta}_{3N+l_4}\; \eta_{3N+l_4-1}
 \;\;\bar{\eta}_{4N+l_5}\; \eta_{4N+l_5} & \cr
 & & \cr
%
%
 & =\cases{ 2^{N-5}, & $l_1 = l_2 + 1 $ and $ l_4 = l_3 +1$, or,   \cr
                     & $ l_1 = l_3 +1$ and $ l_4= l_2 +1$  \cr
2^{N-4}, & $l_1= l_2 + 1$  and $ l_5= l_3 +1$ and $ l_4 = l_5$, or,\cr
         & $l_2 =l_4 =l_1 -1 = l_3 +1$, or,   \cr
         & $l_2= l_5 = l_1 -1$ and $l_3= l_4-1$, or,   \cr
         &  $l_1 = l_3 +1$ and $l_4= l_5= l_2+1$, or,  \cr
         & $l_3 = l_5 = l_1 -1$ and $l_2= l_4 -1$, or,  \cr
         &  $l_1 = l_3 = l_4 -1$ and $l_4= l_2 +2$   \cr
2^{N-3}, & $l_1 =l_3 = l_5+1$ and $l_2 = l_5$ and $l_4 = l_2+2 $, or, \cr
        & $l_1 = l_5 +1$ and $l_2= l_5$ and $l_3= l_5 -1$ and $l_4=l_5$,
              or,   \cr
         & $l_1= l_3 = l_5 -1$ and $ l_2 = l_5 - 2$ and $l_4=l_5$.  \cr}
         & (C.2.4) \cr }$$


\vfill
\eject

\noindent 5)     \par

\vskip -0.5cm

$$\eqalignno{ {\cal G}_5^{(5)} (l_1, l_2, l_3, l_4, l_5)
& \equiv \int \prod_{I=1}^{5N}
\, d\eta_I d\bar{\eta}_I \;\; e^{\sum\limits_{I,J= 1}^{5N}
\bar{\eta}_I\; A_{I J}^{\uparrow \uparrow}
 \; \eta_J} \;\; \bar{\eta}_{l_1} \;\eta_{l_1+1}
\;\;\bar{\eta}_{N+l_2}\; \eta_{N+l_2} \;\; \bar{\eta}_{2N+l_3} \;
\eta_{2N+l_3}  \times  & \cr
& \hskip 3cm
 \times \bar{\eta}_{3N+l_4}\; \eta_{3N+l_4 -1}
 \;\;\bar{\eta}_{4N+l_5}\; \eta_{4N+l_5} & \cr
 & & \cr
%
%
 & =\cases{ 2^{N-5}, & $l_4 = l_1 + 1 $ \cr
2^{N-4}, & $l_4= l_1 + 1$ and two other space indices are equal\cr
2^{N-3}, & $l_4 = l_1+1$ and $l_1 = l_2= l_3 $, or,  \cr
         & all permutations with 3 equal space indices  \cr
         & $l_4 = l_1+1$ and $ l_1 = l_2$ and $l_3= l_5$, or,   \cr
         & all permutations  2 by 2  \cr
2^{N-2}, & $l_4 = l_1 + 1$ and $l_1= l_2=l_3 = l_5-1$.  \cr }
         & (C.2.5) \cr }$$

\vskip 0.3cm


\noindent 6)      \par

\vskip -0.5cm

$$\eqalignno{ {\cal G}_6^{(5)} (l_1, l_2, l_3, l_4, l_5)
& \equiv \int \prod_{I=1}^{5N}
\, d\eta_I d\bar{\eta}_I \;\; e^{\sum\limits_{I,J= 1}^{5N}
\bar{\eta}_I\; A_{I J}^{\uparrow \uparrow}
 \; \eta_J} \;\; \bar{\eta}_{l_1} \;\eta_{l_1+1}
\;\;\bar{\eta}_{N+l_2}\; \eta_{N+l_2-1} \;\; \bar{\eta}_{2N+l_3} \;
\eta_{2N+l_3+1}   \times  & \cr
& \hskip 3cm
 \times \bar{\eta}_{3N+l_4}\; \eta_{3N+l_4 -1}
 \;\;\bar{\eta}_{4N+l_5}\; \eta_{4N+l_5} & \cr
 & & \cr
%
%
 & =\cases{ 2^{N-5}, & $l_2= l_1 + 1 $  and $l_4  =l_3 +1$, or, \cr
                      & $l_4= l_1 +1$ and $ l_2 = l_3+1$    \cr
2^{N-4}, & $l_2= l_1 + 1$ and $l_4= l_5 = l_3 +1$ \cr
         &  $l_2=l_5 = l_1 +1$ and $l_4= l_3 +1$  \cr
         & $ l_4 = l_5= l_1 +1$ and $ l_2= l_3 +1$  \cr
         & $l_1 = l_4 -1$ and $ l_3 = l_5 = l_2 -1$  \cr
2^{N-3}, & $l_1=l_3 = l_4 -1$ and $l_2 = l_4 $\cr
2^{N-2}, & $l_1 = l_3 = l_5 - 1$ and $l_2= l_4=l_5$ \cr }
         & (C.2.6) \cr }$$

\vskip 0.3cm

\noindent 7)    \par

\vskip -0.5cm

$$\eqalignno{ {\cal G}_7^{(5)} (l_1, l_2, l_3, l_4, l_5)
& \equiv \int \prod_{I=1}^{5N}
\, d\eta_I d\bar{\eta}_I \;\; e^{\sum\limits_{I,J= 1}^{5N}
\bar{\eta}_I\; A_{I J}^{\uparrow \uparrow}
 \; \eta_J} \;\; \bar{\eta}_{l_1} \;\eta_{l_1+1}
\;\;\bar{\eta}_{N+l_2}\; \eta_{N+l_2+1} \;\; \bar{\eta}_{2N+l_3} \;
\eta_{2N+l_3-1}  \times  & \cr
& \hskip 3cm
 \times \bar{\eta}_{3N+l_4}\; \eta_{3N+l_4 -1}
 \;\;\bar{\eta}_{4N+l_5}\; \eta_{4N+l_5} & \cr
 & & \cr
%
%
 & =\cases{ 2^{N-5}, & $l_3= l_1 + 1 $  and $l_4 = l_2 +1$, or, \cr
                      & $l_4= l_1 +1$ and $ l_3 = l_2+1$    \cr
2^{N-4}, & $l_3= l_1 + 1$ and $l_4 = l_5 = l_2 +1$, or,  \cr
         &  $l_3 = l_5 = l_1 +1 $ and $l_4= l_2 +1$, or,   \cr
         & $ l_3 = l_5 = l_2 +1$ and $l_4 = l_1 +1$, or,   \cr
         &  $ l_4= l_5= l_1 +1$ and $ l_2 = l_3 -1$, or, \cr
         & $ l_2 = l_4 = l_3 -1 = l_1 +1$, or
               $ l_1 = l_3 = l_4 -1 = l_2 +1 $   \cr
2^{N-3}, & $l_2=l_4 = l_5$ and $l_1 = l_5 -1 $ and $ l_3 = l_5 +1$, or,\cr
          & $ l_1 = l_3 = l_5 -1$ and $l_4 = l_5$ and $ l_2 = l_5 -2$,
               or,   \cr
          & $ l_2 = l_4 =l_5 -1$ and $ l_3 = l_5$ and $l_1= l_5 -2$. \cr}
         & (C.2.7) \cr }$$

\vskip 1cm

\noindent {\bf Acknowledgements} \par

\bigskip

The authors thank J. Florencio Jr. for interesting
discutions and A.T. Costa Jr. for making the figures. 
I.C.C  thanks FAPMG  and E.V.C.S.  thanks CNPq
 for financial support. M.T.T. thanks CNPq and FINEP for
 partial financial support.


\vskip 1cm


\centerline {\bf REFERENCES} \par
\bigskip

\item{1.} {S. Samuel, J. Math. Phys. {\bf 21} (1980) 2806--2833; 
C. Itzykson, Nucl. Phys. {\bf B210} [FS6] (1982) 448; 
V.N. Plechko, Physica A {\bf 152} (1988) 51;
V.N. Plechko and I.K. Sobolev, Physica A {\bf 197} (1993) 323;  }  \par

\item{2.}{I.C. Charret, S.M. de  Souza, E.V. Corr\^ea Silva and  M.T.
 Thomaz, {\it Grand Canonical Partition Function for the Unidimensional
Systems: Application to Hubbard Model Up to Order $\beta^3$},
submitted to  Jour. of  Phys. A,(pre-print cond/mat/ 9607171); }  \par

\item{3. }  { E.H. Lieb and F.Y. Wu, Phys. Rev. Letters {\bf 20} (1968)
1445;  A.A. Ovchinnikov, Sov. Phys. JETP {\bf 30} (1970) 1160;
 }  \par

\item {4.} { M. Takahashi, Prog. Theoret. Phys. (Kyoto) {\bf 43}
(1970) 1619;  }  \par

\item{5.} { K. Kubo and M. Tada, Prog. Theoret. Phys. {\bf 69}
(1983)  1345;
C.J.  Thompson, Y.S. Young, A.J. Guttmann and M.F. Sykes, J. Phys. A:
Math. Gen. {\bf 24} (1991) 1261;
J.A. Henderson, J. Oitmaa and M.C.B. Ashley, Phys. Rev. {\bf B46} (1992)
6328;    }   \par

\item{6.} {  H.E. Stanley, {\it Introduction to Phase Transitions and
Critical Phenomena}, Oxford Univ. Press (1971);}     \par

\item {7.} { C. Itzykson and J.--B. Zuber, {\it Quantum Field
Theory}, McGraw--Hill (1980); \hfill\break U. Wolf, Nucl. Phys.
{\bf B225} [FS9]  (1983) 391;  } \par

\item{8.} {We are using the convention: $\sigma= 1 = \uparrow$
and $\sigma= -1 = \downarrow$; }   \par

\item {9.}{ I.C. Charret, S.M. de  Souza and  M.T. Thomaz,  Braz. Jour.
of Phys. {\bf 26}  (1996) 720;   }  \par

\item{10.}{ J. Hubbard, Proc. Roy. Soc. {\bf A277} (1963) 237; {\bf
A281} (1964) 401;   \hfill \break
M. Gutzwiller, Phys. Rev. Lett. {\bf 10} (1963) 159;
Phys. Rev. {\bf A137} (1965) 1726;} \par

\item{11.}{ I.C. Charret, E.V. Corr\^ea Silva, S.M. de Souza and M.T.
Thomaz, J. Math. Phys. {\bf 36} (1995) 4100;}   \par

\item{12.} {See section 4.1 of reference [2] for a discussion on 
the vanishing Grassmann integrals;}   \par

\item{13.} { I.S. Gradshteyn and I.M. Ryzhik; {\it Table of Integrals,
Series and Products}, $4^{th}$ edition, Academic Press (1965);
expression: 1.396.4;}  \par

\item{14.} { Reference [11], expression: 1.392.2;}  \par

\item{15.} { When the space index does not appear among the conditions,
we mean that it is distinct from any other space index in the
graph;}  \par

\item{16.} { O.R. dos Santos and M.T. Thomaz, private communication.}

\vfill
\eject
\bye